\documentclass[draftcls,12pt,onecolumn]{IEEEtran}
\usepackage{stfloats,color}
\usepackage{amsfonts}
\usepackage[cmex10]{amsmath}
\usepackage{graphicx}
\usepackage{setspace}
\usepackage{cite}
\usepackage{array}
\usepackage{algorithmic}
\usepackage{mdwmath}
\usepackage{mdwtab}
\usepackage[center]{caption}
\usepackage{fixltx2e}
\usepackage{url}
\usepackage{times}
\usepackage{epsfig}
\usepackage{latexsym}
\usepackage{epstopdf}
\usepackage{verbatim}
\usepackage{units}
\usepackage{amsthm}
\usepackage{mdwlist}
\usepackage{multicol}
\usepackage{paralist}
\usepackage{subfig}
\allowdisplaybreaks
\usepackage{soul}\graphicspath{{./figures/}}

\begin{document}


\title{Stability Analysis of Slotted Aloha with Opportunistic RF Energy Harvesting}


\author{Abdelrahman~M.Ibrahim, Ozgur~Ercetin, and~Tamer~ElBatt
\thanks{Abdelrahman~M.Ibrahim is with the Department of Electrical Engineering, The Pennsylvania State University, University Park, PA 16802. This work was done when he was with the Wireless Intelligent Networks Center (WINC), Nile University, Giza, Egypt (e-mail: ami137@psu.edu).}%
\thanks{Ozgur~Ercetin is with the Faculty of Engineering and Natural Sciences, Sabanci University, Istanbul, Turkey.}%
\thanks{Tamer~ElBatt is with the Wireless Intelligent Networks Center (WINC), Nile University, Giza, Egypt. He is also affiliated with the EECE Dept., Faculty of Engineering, Cairo University, Egypt.}
\thanks{This material is based upon work supported by the Marie Curie International Research Staff Exchange Scheme Fellowship PIRSES-GA-2010-269132 AGILENet within the 7th European Community Framework Program.}
}
\maketitle

\begin{abstract}
 Energy harvesting (EH) is a promising technology for realizing energy efficient wireless networks. In this paper, we utilize the ambient RF energy, particularly interference from neighboring transmissions, to replenish the batteries of the EH enabled nodes. However, RF energy harvesting imposes new challenges into the analysis of wireless networks. Our objective in this work is to investigate the performance of a slotted Aloha random access wireless network consisting of two types of nodes, namely Type I which has unlimited energy supply and Type II which is solely powered by an RF energy harvesting circuit. The transmissions of a Type I node are recycled by a Type II node to replenish its battery. We characterize an inner bound on the stable throughput region under half-duplex and full-duplex energy harvesting paradigms as well as for the finite capacity battery case. We present numerical results that validate our analytical results, and demonstrate their utility for the analysis of the exact system. 




\end{abstract}



\section{Introduction} \label{sec:Introduction}
One of the prominent challenges in the field of communication networks today is the design of energy efficient systems. In traditional networks, wireless nodes are powered by limited capacity batteries which should be regularly charged or replaced. Energy harvesting has been recognized as a promising solution to replenish batteries without using any physical connections for charging. Nodes may harvest energy through solar cells, piezoelectric devices, RF signals, etc. In this paper, we focus on RF energy harvesting. Recent studies present experimental measurements for the amount of RF energy that can be harvested from various RF energy sources. Two main factors affect the amount of RF energy that can be harvested, namely, the frequency of the RF signal and the distance between the ``interferer" and the harvesting node, e.g., see Table I in \cite{lu2014dynamic}.\\ 
\indent Recently, an information-theoretic study of the capacity of an additive white Gaussian noise (AWGN) channel with stochastic energy harvesting at the transmitter has shown that it is equal to the capacity of an AWGN channel under an average power constraint \cite{ozel2012achieving}. This, in turn, motivated the investigation of optimal transmission policies \cite{gunduz2014designing} for single user \cite{yang2012optimal,tutuncuoglu2012optimum,ozel2011transmission} and multi-user \cite{yang2012broadcasting,yang2012ma,tutuncuoglu2012sum} energy harvesting networks. The optimal policy that minimizes the transmission completion time was studied in \cite{yang2012optimal}. In \cite{tutuncuoglu2012optimum}, the authors studied the problem of maximizing the short-term throughput and have shown that it is closely related to the transmission completion time problem \cite{yang2012optimal}. The authors in \cite{ozel2011transmission} studied the optimal transmission policies for energy harvesting networks under fading channels. Moreover,  \cite{yang2012broadcasting,yang2012ma,tutuncuoglu2012sum} extends the analysis to broadcast, multiple access, and interference channels, respectively. The authors in \cite{gurakan2012energy} introduced the concept of energy cooperation where a user can transfer portion of its energy, over a separate channel, to assist other users.\\ 
\indent Significant research has also been conducted on RF energy harvesting. In \cite{varshney2008transporting}, the author discusses the fundamental trade-offs between transmitting energy and transmitting information over a single noisy link. The author derives the capacity-energy functions for several channels. The authors in \cite{fouladgar2012transfer}, extend the point-to-point results of \cite{varshney2008transporting} to multiple access and multi-hop channels. Recently, several techniques were proposed for designing RF energy harvesting networks (RF-EHNs), e.g., \cite{lu2014survey}. The RF energy harvesting process can be classified as follows: \begin{inparaenum}[\itshape i\upshape)]
\item \emph{Wireless energy transfer}, where the transmitted signals by the RF source are dedicated to energy transfer,
\item \emph{Simultaneous wireless information and power transfer}, where the transmitted RF signal is utilized for both information decoding and RF energy harvesting, and
\item \emph{Opportunistic energy harvesting}, where the ambient RF signals,  considered as interference for data transmission, are utilized for RF energy harvesting.  
\end{inparaenum}
The receiver architecture may also vary as follows \cite{lu2014survey},\cite{zhang2011mimo}: \begin{inparaenum}[\itshape a\upshape)]
\item \emph{Co-located receiver architecture}, where the radio receiver and the harvesting circuit use the same antenna for both decoding the data and energy harvesting, and
\item \emph{Separated receiver architecture}, where the radio receiver and the harvesting circuit are separated, and each is equipped with its own antenna and RF front-end circuitry.
\end{inparaenum}
\newline \indent In \cite{zhang2011mimo}, the authors discuss the practical limitations of implementing a simultaneous wireless information and power transfer (SWIPT) system. A major issue is that energy harvesting circuits are not able to simultaneously decode information and harvest energy. Hence, the authors in \cite{zhang2011mimo} proposed and analyzed two modes of operation for the co-located receiver architecture, that is, time switching and power splitting. Furthermore, the RF energy harvesting transceivers may also be classified as: \begin{inparaenum}[\itshape a\upshape)]
\item \emph{Half-duplex energy harvester}, where a co-located RF energy harvesting transceiver can either transmit data or harvest RF signals at a given instant of time, and
\item \emph{Full-duplex energy harvester}, where a node is equipped with two independent antennas and can transmit data and harvest RF signals, simultaneously. In this paper, we investigate opportunistic RF energy harvesting under the half-duplex and full-duplex modes of operation.
\end{inparaenum}
\newline \indent The cornerstone of random access medium access control (MAC) protocols is the Aloha protocol \cite{abramson1970aloha}, which is widely studied in multiple access communication systems because of its simplicity. The applications of Aloha-based protocols range from traditional satellite networks \cite{okada1977analysis} to radio frequency identification (RFID) systems \cite{zhu2011critical} and the emerging Machine-to Machine (M2M) communications \cite{wu2013fasa}. It is also considered as a benchmark for evaluating the performance of more sophisticated MAC protocols. Based on the Aloha protocol, nodes contend for the shared wireless medium and cause interference to each other. Hence, the service rate of a node depends on the backlog of other nodes, \.i.e., the nodes' queues become \emph{interacting} as originally characterized in \cite{fayolle1979two}. Tsybakov and Mikhailov \cite{tsybakov1979ergodicity} were the first to analyze the stability of a slotted Aloha system with finite number of users. Rao and Ephremides \cite{rao1988stability} characterized the sufficient and necessary conditions for queue stability of the two user case, using the so-called \emph{stochastic dominance} technique. In addition, they established conditions for the stability of the symmetric multi-user case. Other works followed and studied the stability of slotted Aloha with more than two users \cite{luo1999stability, bordenave2012asymptotic,kompalli2013stability,szpankowski}. The authors in \cite{ghez1988stability} extended the stability analysis under the collision model to a symmetric multi-packet reception (MPR) model, which was later generalized to the asymmetric MPR model in \cite{naware2005stability}.\\ 
\indent Perhaps the closest to our work are \cite{jeon2011stability,jeon2015stability} which characterize the stability region of a slotted Aloha system with energy harvesting capabilities, under the multi-packet reception model. The authors considered a system where the nodes harvest energy from the environment at a fixed rate and, thus, the energy harvesting process is modeled as a Bernoulli process.\\
\indent In this work, we analyze the stability of a slotted Aloha random access wireless network consisting of nodes with and without RF energy harvesting capability. Specifically, we consider a wireless network consisting of two nodes, namely a node of Type I which has unlimited energy supply and a node of Type II which is powered by an RF energy harvesting circuit. The RF transmissions of the Type I node are harvested by the Type II node to replenish its battery. Our contribution in this paper is multi-fold. First, we outline the difficulties in analyzing the stability of the exact RF energy harvesting Aloha system $\mathcal{S}_O$ and for mathematical tractability we introduce an equivalent system $\mathcal{S}_G$. Second, we generalize the \emph{stochastic dominance technique} for analyzing RF EH-networks. Third, we characterize an inner bound on the stable throughput region of the system $\mathcal{S}_G$ under the half-duplex and full-duplex energy harvesting paradigms. Also, we derive 
the closure of the inner bound over all transmission probability vectors. Fourth, we investigate the impact of finite capacity batteries on the stable throughput region. Finally, we validate our analytical findings with simulations and conjecture that the inner bound of the system $\mathcal{S}_G$ is also an inner bound for the exact system $ \mathcal{S}_O$.
\newline \indent The rest of this paper is organized as follows. In Section \ref{sec:System_Model}, we present the system model and the assumptions underlying our analysis. In Section \ref{sec:Energy_Harvesting_Model}, we describe the energy harvesting models for the systems $\mathcal{S}_O $ and $\mathcal{S}_G$. Our main results are presented in Section \ref{sec:Main_Results} and proved in Section \ref{sec:Stability_Analysis}. In Section \ref{sec:Extensions}, we investigate the impact of finite capacity batteries and full-duplex energy harvesting on the stability region of our system. We corroborate our analytical findings by simulating the systems $\mathcal{S}_O $ and $\mathcal{S}_G$ in Section \ref{sec: Numerical}. Finally, we draw our conclusions and point out directions for future research in Section \ref{sec:conclusions}.
%
\section{System Model} \label{sec:System_Model}
\begin{figure}[t]
\includegraphics[width=0.4\textwidth, height=0.2\textheight]{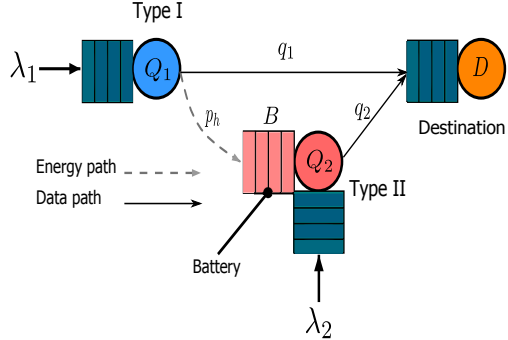}
\centering
\caption{System model}\label{fig:EH_model}
\end{figure}

We consider a wireless network consisting of two source nodes and a common destination, as shown in Fig. \ref{fig:EH_model}. We consider a slotted Aloha multiple access channel \cite{abramson1970aloha}, where time is slotted and the slot duration is equal to one packet transmission time. We assume two types of nodes in our system: Type I node has a data queue, $Q_1$, and unlimited energy supply, while Type II node has a data queue, $Q_2$, and a battery queue, $B$, as shown in Fig. \ref{fig:EH_model}. Moreover, packets arrive to the data queues, $ Q_1$ and $Q_2$, according to independent Bernoulli processes with rates $\lambda_1$ and $\lambda_2$, respectively. The transmission probabilities of Type I and II nodes are $q_1$ and $q_2$, respectively.\\
\indent We assume perfect data channels, \.i.e., the destination successfully decodes a data packet, if only one node transmits. If two nodes transmit simultaneously, a collision occurs and both packets are lost and have to be retransmitted in future slots. At the end of each time slot, the destination sends an immediate acknowledgment (ACK) via an error-free feedback channel.

Data packets of Type I and II nodes are stored in queues $Q_1$ and $Q_2$, respectively. The evolution of the queue lengths is given by \cite{szpankowski} 
\begin{equation}\label{eqn: q evl}
Q_i^{t+1}=\max\lbrace Q_i^{t}-Y_i^{t},0\rbrace+ X_i^{t}, \ \ \ i=1,2.
\end{equation}
where $X_i^{t} \in \lbrace 0,1 \rbrace$ is the arrival process for data packets and $Y_i^{t} \in \lbrace 0,1 \rbrace$ is the departure process independent of $Q_i$ status, \.i.e., $Y_i^{t}=1$ even if the data queue is empty \cite{naware2005stability}. $X_i^{t}$ is a Bernoulli process with rate $\lambda_i$ and $\mathbb{E}[ Y_i^{t}]=\mu_i$. 

We assume that Type II nodes operate under a half-duplex energy harvesting mode, \.i.e., they either harvest or transmit but not both simultaneously\footnote{We extend the analysis to the case of full-duplex RF EH in Section \ref{sec:Extensions}.}. Hence, the harvesting opportunities are in those slots when a Type II node is idle while a Type I node is transmitting. The channel between the two source nodes is a block fading channel, where the fading coefficient remains constant within a single time slot and changes independently from a slot to another. For Rayleigh fading, the instantaneous channel power gain $h_t$ at time slot $t$ is exponentially distributed, \.i.e., $h_t \sim \textit{Exp}(1)$. Let $P_j$ be the transmission power of node $j$, and $l$ be the distance between the two source nodes. We consider the non-singular\footnote{In order to harvest significant amount of RF energy, $l$ is typically small. Hence, we use a non-singular (bounded) pathloss model instead of a singular (unbounded) pathloss $l^{-\alpha}$ model, because the singular pathloss model is not correct for small values of $l$ due to singularity at $0$, \cite{2009unbounded}. } pathloss model with $(1+l^\alpha)^{-1}$, where $\alpha$ is the pathloss exponent. We use power and energy interchangeably throughout the paper, since we assume unit time slots. In the next section, we develop a discrete-time stochastic process to model RF energy harvesting. 
%
%

\section{Energy Harvesting Models} \label{sec:Energy_Harvesting_Model}

Prior work largely models the energy harvesting process as an independent and identically distributed (iid) Bernoulli process with a constant rate \cite{jeon2015stability}\footnote{Typically, the energy harvesting process is not iid, because to harvest one energy unit, energy is accumulated over multiple slots.}. In the following, we first study the RF energy harvesting process of the exact system $\mathcal{S}_O$ and show that the inter-arrival time of the energy arrivals has a general distribution. Second, we propose an equivalent system $\mathcal{S}_G$ where the inter-arrival time of the energy arrivals is geometrically distributed with the same mean as $\mathcal{S}_O$. 

%
\vspace{-0.2in}
\subsection{RF Energy Harvesting Model of the exact system $\mathcal{S}_O$}\label{sec: exact system}
\indent The received power at a Type II node from the transmissions of Type I node at time slot $t$ is $ P_{R}(t)= \eta P_1 h_t (1+l^\alpha)^{-1}$, where $P_{1}$ is the transmission power of Type I node and $\eta$ is the RF harvesting efficiency \cite{lu2014survey}. Recent studies demonstrated that $\eta$ typically ranges from $0.5$ to $0.7$, where its value depends on the efficiency of the harvesting antenna, impedance matching circuit and the voltage multiplier \cite{coporation2011rf}. In order to develop the analytical model underlying this paper, we approximate the continuous energy arrival process in quantas of size $\gamma$ joules.  

\begin{figure}[t]
\includegraphics[width=0.42\textwidth, height=0.24\textheight]{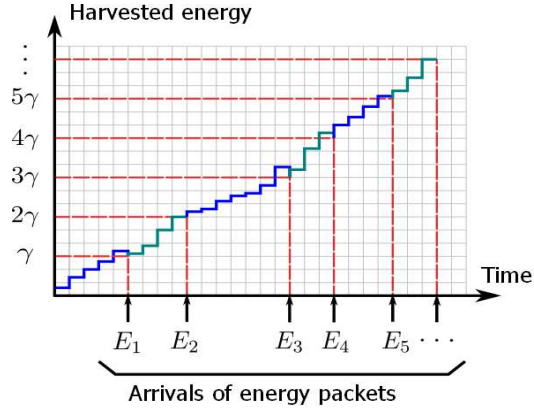}
\centering
\caption{A unit of energy is harvested every time the accumulated energy in the battery exceeds $\gamma$.}\label{fig:energy}
\end{figure}
\indent Typically, we need to harvest RF energy from multiple transmissions of Type I node in order to accumulate $\gamma$ joules. Conceptually, accumulating $\gamma$ joules is equivalent to having a single {\em energy packet arrival} to the battery. We model the battery of a Type II node as a queue with unit energy packet arrivals from the harvesting process, and unit energy packet departures when Type II node transmits. There is an energy packet arrival to the battery queue at the end of the time slot in which the accumulated energy exceeds $\gamma$, see Fig. \ref{fig:energy}. Let $Z$ be the number of Type I transmissions needed to harvest one energy unit. 

\indent \textbf{Lemma 1.} For a persistent ($q_1\!=\!1$) and saturated Type I node, the probability mass function (PMF) of $Z$, when the channel between Type I and II nodes is modeled as a Rayleigh fading channel with parameter $h_t$, is given by
\begin{align}
\mathbb{P}[Z=k]&=\frac{e^{-\theta} \ \theta^{k-1}}{(k-1)!}, \ k=1,2,\cdots, \ \text{where } \theta=\frac{\gamma (1+l^\alpha)}{\eta P_1}
\end{align}
\begin{proof}
Note that if the accumulated received energy over $k$ slots is greater than or equal to, $\gamma$ while the accumulated received energy up to slot $k-1$ is less than $\gamma$, then we need $k$ slots to harvest one energy unit.
\begin{align} 
\mathbb{P}[Z=k]&= \mathbb{P}\left[  \sum_{t=1}^{k} P_{R}(t) \geq \gamma, \ \sum_{t=1}^{k-1} P_{R}(t) < \gamma\right]  \\
&\stackrel{(a)}{=} \mathbb{P}\left[ \sum_{t=1}^{k-1} h_{t} < \theta \right]- \mathbb{P}\left[ \sum_{t=1}^{k} h_{t} < \theta\right] \label{egn: slot dist 1} \\
&\stackrel{(b)}{=} \frac{e^{-\theta} \ \theta^{k-1}}{(k-1)!}, \ k=1,2,\cdots, \label{egn: slot dist 2}
\end{align}
(a) follows from applying the law of total probability, \.i.e, 
\begin{align}
\mathbb{P}\left[ \sum_{t=1}^{k-1} h_{t} < \theta \right] &= \mathbb{P}\left[ \sum_{t=1}^{k-1} h_{t} < \theta, \sum_{t=1}^{k} h_{t} \geq \theta \right] + \mathbb{P}\left[ \sum_{t=1}^{k-1} h_{t} < \theta, \sum_{t=1}^{k} h_{t} < \theta \right],\\
&= \mathbb{P}\left[ \sum_{t=1}^{k-1} h_{t} < \theta, \sum_{t=1}^{k} h_{t} \geq \theta \right] + \mathbb{P}\left[ \sum_{t=1}^{k} h_{t} < \theta \right].
\end{align}  
The distribution of sum of independent exponential random variables is an Erlang Distribution. Hence, $\sum_{t=1}^{k} h_{t} \sim \textit{Erlang}(k,1)$, where $k$ is the shape parameter and $h_t \sim \textit{Exp}(1)$. From, the cumulative distribution function (CDF) of the Erlang distribution, we know that   
\begin{align} \label{eqn: cdf Erlang}
\mathbb{P}\left[ \sum_{t=1}^{k} h_{t} < \theta \right]= 1- \sum_{j=0}^{k-1} \frac{e^{-\theta} \ \theta^{j}}{j!}, 
\end{align}
Hence, (b) is obtained by substituting (\ref{eqn: cdf Erlang}) in (\ref{egn: slot dist 1}). 
\end{proof}
From (\ref{egn: slot dist 2}), we notice that the distribution of the inter-arrival times of the energy harvesting process is a shifted Poisson distribution, \.i.e., $Z=V+1$, where $V$ is a Poisson random variable with mean $\theta$. The expected inter-arrival time of the harvesting process is given by $\mathbb{E}[Z]=1+\theta$. In the general case where Type I node is unsaturated and transmits with probability $q_1$, characterizing the PMF of $Z$ is challenging because the queue evolution process is not an iid process. For instance, if $Q_1^t=2$, we know for sure that $Q_1^{t+1}\geq 1$.



\subsection{RF Energy Harvesting Model of the equivalent system $\mathcal{S}_G$}

For the mathematical tractability of the results derived in later sections, we consider a system $\mathcal{S}_G$, where the RF energy harvesting process is an iid Bernoulli process. Let $p_{h|\lbrace 1 \rbrace}$ be the mean of the iid Bernoulli process of unit energy packet arrivals, where it can be interpreted as the probability of success in harvesting one energy unit given that Type I node is transmitting. Based on the fact that the inter-arrival time of a Bernoulli process is geometrically distributed, the mean inter-arrival time is $1/p_{h|\lbrace 1 \rbrace}$. Hence, the relationship between the exact harvesting process in $\mathcal{S}_O$ and the equivalent Bernoulli process in $\mathcal{S}_G$ is given by
\begin{equation}\label{eqn: theta}
p_{h|\lbrace 1 \rbrace} = \frac{1}{1+\theta}.
\end{equation}
\indent In general, an iid Bernoulli process has a rate $p_{h|\mathcal{M}}$, where $p_{h|\mathcal{M}}$ is the probability of harvesting one energy unit given a set of nodes $\mathcal{M}$ are transmitting. Under half-duplex energy harvesting, Type II node only harvests from the transmissions of Type I node, when the Type II node is not transmitting, \.i.e., the probability of harvesting one energy unit given Type II node is transmitting $p_{h|\lbrace 2 \rbrace}=0$, and the probability of harvesting one energy unit given both nodes are transmitting $p_{h|\lbrace 1,2 \rbrace}=0$. For convenience, we denote $p_{h|\lbrace 1 \rbrace}$ by $p_{h}$. 



%

%
\subsection{Analyzing the Battery queue in $\mathcal{S}_G$}
\begin{figure}
\includegraphics[width=0.55\textwidth, height=0.1\textheight]{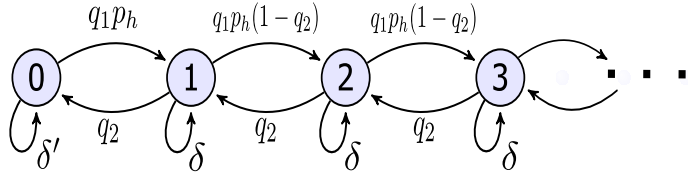}
\centering
\caption{Markov chain model of the battery queue given that the data queues $Q_1$ and $Q_2$ are backlogged. Note that 
$\delta^{\prime}\! = \! 1\!- \!q_1 p_h \! - q_2$ and $ \delta \! = \! 1\!- \!q_1 p_h (1-q_2) \! - q_2$.}\label{fig:MC_half_duplex}
\end{figure}

We assume that a Type II node opportunistically harvests RF energy packets from the transmissions of Type I node. Also, transmitting a single data packet costs one energy unit. Let $H^t$ denotes the energy harvesting process modeled as a Bernoulli process. Assuming half-duplex harvesting,\footnote{In the sequel, we will extend the model to full-duplex as well.} the average harvesting rate is the difference between the fraction of time slots in which the Type I node is transmitting and the fraction of time slots in which both nodes are transmitting, \.i.e.,
\vspace{-0.2in}
 \begin{equation}\label{eqn: harvt rate}
 \mathbb{E}[H^{t}] = q_1 p_{h} \ \mathbb{P}[Q_1 \! > \! 0]-q_1 q_2 p_{h} \ \mathbb{P}[Q_1 \! > \! 0, B \! > \!0, Q_2 \! > \! 0]. 
 \end{equation}
The battery queue evolves as \cite{jeon2015stability}
\vspace{-0.2in}
\begin{equation}
B^{t+1}=B^{t}-\mu_B^{t} + H^{t},
 \end{equation}
where $\mu_B^{t} \in \lbrace 0,1 \rbrace$ represents the energy consumed in the transmission of a data packet at time $t$.
Under backlogged data queues $Q_1$ and $Q_2$, the average rate of harvesting becomes 
\begin{equation}
 \mathbb{E}[H^{t} | Q_1 \! > \! 0, Q_2 \! > \!0] = q_1 p_{h} \big ( 1-q_2 \mathbb{P}[B \! > \!0| Q_1 \! > \! 0, Q_2 \! > \!0] \big ). 
 \end{equation}
\indent Now, the energy harvesting rate is only dependent on the battery queue status. Thus, if the battery queue is empty, the energy harvesting rate is $q_1 p_h$, otherwise it is reduced to $q_1 p_h (1-q_2)$ because of the half-duplex operation. Hence, the battery queue forms a decoupled discrete-time Markov chain, as shown in Fig. \ref{fig:MC_half_duplex}. By analyzing the Markov chain we find the probability that the battery is non-empty, as given by the following lemma. \newline
\indent \textbf{Lemma 2.} For a half-duplex RF energy harvesting node with infinite capacity battery, the probability that the battery is non-empty, given that the data queues $Q_1$ and $Q_2$ are backlogged, is given by 
\begin{equation}
\mathbb{P}[B>0| Q_1>0, Q_2 >0]= \min \left\lbrace \frac{q_1 p_{h}}{q_2 (1+ q_1 p_{h})}, 1\right\rbrace .
\end{equation}
\begin{proof}
Let $\boldsymbol{\pi}=[\pi_0,\pi_1,...]$ be the steady-state distribution of the Markov chain shown in Fig. \ref{fig:MC_half_duplex}. Applying the detailed balance equations, we obtain 
$ \pi_k = \left( \frac{q_1 p_{h}}{q_2}\right)^k (1-q_2)^{k-1} \pi_0, \ k=1,2,\cdots. $
Therefore, by substituting in the normalization condition $\sum_i \pi_i =1$, we get the utilization factor 
$ \rho= 1-\pi_0 = \frac{q_1 p_{h}}{q_2(1+q_1 p_{h})}.  $
Hence,  $ \mathbb{P}[B>0| Q_1>0, Q_2>0]= \min \left\lbrace \rho, 1\right\rbrace. $
\end{proof}

\vspace{-0.2in}
%
\section{Main Results} \label{sec:Main_Results}
In this section, we present our main results pertaining to the stable throughput region of 
the opportunistic RF energy harvesting slotted Aloha network $\mathcal{S}_G$. We adopt the notion of stability proposed in \cite{szpankowski}, where the stability of a queue is determined by the existence of a proper limiting distribution. A queue is said to be stable if 
\begin{equation}  
\lim_{t \rightarrow \infty} \mathbb{P}[Q^t < x]=F(x) \textit{ and } \lim_{x \rightarrow \infty} F(x)=1.
\end{equation}
\indent The stability of a queue is equivalent to the recurrence of the Markov chain modeling the queue length. \emph{Loynes' Theorem} \cite{loynes} states that if the arrival and service processes of a queue are strictly jointly stationary and the average arrival rate is less than the average service rate, then the queue is stable. Also, if the average arrival rate is greater than the average service rate, then the queue is unstable and the queue size $Q^t$ approaches infinity almost surely. The stable throughput region of a system is defined as the set of arrival rate vectors, $(\lambda_1,\lambda_2)$ for our system, for which all data queues in the system are stable.\\
\indent Next, we establish sufficient conditions on the stability of the opportunistic RF energy harvesting Aloha $\mathcal{S}_G$. Assuming half-duplex energy harvesting and unlimited battery capacity, the stability region is characterized by the following theorem.\\
\indent \textbf{Theorem 1.} An inner bound on the stable throughput region of the opportunistic RF energy harvesting slotted Aloha $\mathcal{S}_G$ is the triangle OBD, shown in Fig. \ref{fig:region}. Assuming half-duplex energy harvesting and unlimited battery size, the region is characterized by  
\begin{align} \label{eqn thm1}
\mathbb{R}_G^{\text{inner}}= \bigg\{  (\lambda_1, \lambda_2) \ \big | \ \lambda_1 \leq  q_1 \left( 1- \frac{\lambda_2}{1-q_1}\right),  \lambda_2 \leq  \dfrac{(1-q_1) \min \left\lbrace \frac{q_1 p_h}{ (1+ q_1 p_h)}, q_2 \right\rbrace \lambda_1}{q_1 \left( 1- \min \left\lbrace \frac{q_1 p_h}{ (1+ q_1 p_h)}, q_2 \right\rbrace \right)} \bigg\}. 
\end{align}
\begin{proof}
The proof is established in the Sections \ref{subsec: Ser G} to \ref{subsec: rel}.
\end{proof}
\indent \textbf{Theorem 2.} The closure of the inner bound $\mathbb{R}_G^{\text{inner}}$ over all transmission probability vectors $\boldsymbol{q}=(q_1,q_2)$ is characterized by 
 \begin{align} \label{eqn: closure}
\mathcal{C}_{(\mathbb{R}_G^{\text{inner}})}= \bigcup\limits_{\boldsymbol{q}}  \ \mathbb{S}_o(\boldsymbol{q})=\bigg\{ \! (\lambda_1, \lambda_2) \ \big |  \lambda_2 \leq \frac{p_h \lambda_1}{2} \! \left( 1 \! - \! \lambda_1 \! + \! \lambda_2 \! + \! \sqrt{(1 \! + \! \lambda_1 \! - \! \lambda_2)^2 \! - \! 4 \lambda_1}\right) \! \! \bigg\}.
\end{align}
\begin{proof}
The proof is established in Section \ref{subsec: closure}.
\end{proof}

%
%
\section{Stability Analysis} \label{sec:Stability_Analysis}
For the majority of prior work on stability analysis of interacting queues, the service rate of a typical node decreases with respect to the transmissions of other nodes in the system. Perhaps, the most basic example is the conventional slotted Aloha system \cite{rao1988stability}, where increasing the service rate of an arbitrary node comes at the expense of decreasing the service rate of other nodes. For our purposes, we refer to such systems without energy limitations as ``interference-limited" systems.\\ 
\indent On the other hand, in our RF energy harvesting system, transmissions from interfering nodes give rise to two opposing effects on Type II (RF energy harvesting) nodes. Similar to classic interference-limited systems, the interfering nodes create collisions and, thus, decrease the service rate of RF energy harvesting nodes. Meanwhile, transmissions from interfering nodes are exploited by RF energy harvesting nodes to opportunistically replenish their batteries. Therefore, from the perspective of an RF energy harvesting node, a fundamental trade-off prevails between the increased number of energy harvesting opportunities and the increased collision rate, which are both caused by interference. As will be shown formally, this fundamental trade-off splits the stable throughput region for RF energy harvesting slotted Aloha networks into two sub-regions, a sub-region where interference is advantageous for the RF energy harvesting node and another sub-region where it is not. These two sub-regions map directly to two modes of operation for our system and are characterized as follows:
\begin{enumerate}
\item \emph{Energy-limited mode:} is the sub-region of the stable throughput region in which the transmissions of interfering nodes \emph{enhance} the throughput of the RF energy harvesting node, \.i.e, the throughput enhancement due to the increased harvesting opportunities outweights the degradation due to collisions created by the interfering nodes.
\item \emph{Interference-limited mode:} is the sub-region of the stable throughput region in which the transmissions of interfering nodes \emph{do not increase} the throughput of the RF energy harvesting node, \.i.e, the throughput degradation due to collisions equals or outweighs the throughput increase due to the increased energy harvesting opportunities.
\end{enumerate} 
\begin{figure}
\includegraphics[width=0.45\textwidth, height=0.26\textheight]{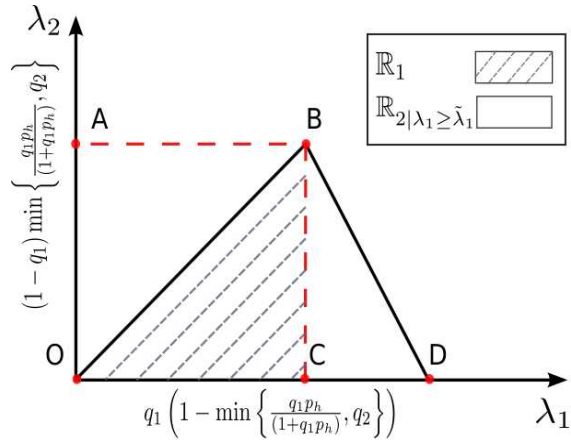}
\centering
\caption{The energy limited and interference limited sub-regions of the stable throughput region of $\mathcal{S}_D$ are characterized by the triangles BCO and BCD, respectively.}\label{fig:region}
\end{figure}
\indent The two parts of the stable throughput region for our system are shown in Fig. \ref{fig:region}, where the energy-limited region is enclosed by the triangle BCO and the interference-limited region is enclosed by the triangle BCD.\\
\indent In order to characterize the inner bound on the stability region of $\mathcal{S}_G$ in Theorem 1, we introduce a deprived system $\mathcal{S}_D$ where Type II node transmits only in the time slots in which $Q_1$ is non-empty. We derive the stability region of $\mathcal{S}_D$, by going through the following three steps discussed next. First, we characterize the average service rates of the two interacting data queues. Second, we generalize the \emph{Stochastic dominance technique} proposed in \cite{rao1988stability} to capture our system dynamics and two-mode operation. Third, we derive the stability conditions of $\mathcal{S}_D$ using the generalized stochastic dominance approach. Finally, we prove in Section \ref{subsec: rel} that the stability region of $\mathcal{S}_D$ is an inner bound on the stability region of $\mathcal{S}_G$.
\vspace{-0.15in}
%
\subsection{Service Rates of the Interacting Queues in $\mathcal{S}_G$}\label{subsec: Ser G}
The average service rates of the data queues, $Q_1$ and $Q_2$, in $\mathcal{S}_G$ are given by 
\begin{align}
\mu_1 &= q_1 \big(1-q_2 \ \mathbb{P}[B \! > \!0, Q_2 \!> \!0| Q_1 \! > \!0] \big), \label{eqn: service1 G}\\
\mu_2 &= q_2 \ \mathbb{P}[B\!>\!0, Q_1\!=\! 0| Q_2\!>\!0]  + q_2 \left( 1-q_1\right) \ \mathbb{P}[B\!>\!0, Q_1\!>\! 0| Q_2\!>\!0]. \label{eqn: service2 G}
\end{align}
where the service rate of the Type I node, $\mu_1$, is the fraction of time in which Type I node decides to transmit, excluding the fraction of time in which Type II node is also transmitting. A Type II node transmits, if it is active, \.i.e., $B \! > \!0$ and $Q_2 \! > \!0$, and decides to transmit. Similarly, the service rate of the Type II node, $\mu_2$, is the fraction of time in which Type II node has a non-empty battery and decides to transmit, excluding the fraction of time in which both nodes are transmitting. Note that the queue evolution equation in (\ref{eqn: q evl}) implies that $ \mathbb{P}[Q_1>0]=1$ in (\ref{eqn: service1 G}) and $ \mathbb{P}[Q_2>0]=1$ in (\ref{eqn: service2 G}). 

In our system, we have three interacting queues, namely $Q_1$, $Q_2$ and $B$. The analysis of three interacting queues is prohibitive and, hence, calculating the probability $ \mathbb{P}[B\!>\!0, Q_1\!=\! 0| Q_2\!>\!0]$. Therefore, we consider a system $\mathcal{S}_D$ where $ \mathbb{P}[B\!>\!0, Q_1\!=\! 0| Q_2\!>\!0 ]\!=\!0$, \.i.e., we consider a lower service rate for the Type II node, since it transmits only in the time slots in which $Q_1$ is non-empty. The relationship between the two systems is discussed in Section \ref{subsec: rel}.
\vspace{-0.15in}
%
\subsection{Service Rates of the Interacting Queues in $\mathcal{S}_D$}
In order to analyze the interaction between $Q_1$ and $Q_2$ in $\mathcal{S}_D$, we decouple the battery queue, $B$, by substituting the probability of the battery queue being non-empty with the conditional probability given by Lemma 2. Hence, the average service rates of the data queues $Q_1$ and $Q_2$ are given as 
\vspace{-0.4in}
\begin{align}
\mu_1 &= q_1 \big(1-q_2 \ \mathbb{P}[B\!>\!0| Q_1\!>\! 0, Q_2\!>\!0]  \ \mathbb{P}[ Q_2\!>\!0| Q_1\!> \!0] \big) \label{eqn: service1 D}\\
&=  q_1 \big( 1- \min \big\lbrace \frac{q_1 p_h}{ (1+ q_1 p_h)}, q_2 \big\rbrace \ \mathbb{P}[ Q_2\!>\!0| Q_1\!> \!0] \big). \label{eqn: int mu1} \\
\mu_2 &= q_2 \left( 1-q_1\right)  \mathbb{P}[B\!>\!0| Q_1\!>\! 0, Q_2\!>\!0]  \ \mathbb{P}[ Q_1\!>\!0| Q_2\!> \!0]   \label{eqn: service2 D}\\
&= (1-q_1) \min \big\lbrace \frac{q_1 p_h}{ (1+ q_1 p_h)}, q_2 \big\rbrace  \ \mathbb{P}[ Q_1\!>\!0| Q_2\!> \!0]. \label{eqn: int mu2}
\end{align}
The probability that $Q_i$ is non-empty given that $Q_j$ is saturated (always backlogged) is given by $\mathbb{P}[ Q_i >0| Q_j > 0]=\frac{\lambda_i}{\mu_i^s}, \ i=1,2, \ i\neq j $, where $\mu_i^s$ is the service rate of $Q_i$ given that both data queues are saturated. We derive $\mu_i^s, \ i=1,2$ in Section \ref{subsec: conditions}.\\
\indent From (\ref{eqn: int mu1}), we note that $\mu_1$ decreases with increasing $\mathbb{P}[ Q_2 >0| Q_1 > 0]$. Recall that for the interference-limited region, increasing the service rate of one node comes at the expense of decreasing the service rate of other nodes. The system is interference-limited from the perspective of Type I node, since increasing $\lambda_2$ comes at the expense of decreasing $\mu_1$. Also, from (\ref{eqn: int mu2}), we observe that $\mu_2$ is increasing in $\mathbb{P}[ Q_1 >0| Q_2 > 0]$.
Hence, increasing $\lambda_1$ increases $\mu_2$ until both data queues are saturated. Thus, the system is energy-limited from the perspective of the Type II node until the data queues become saturated.

In Fig. \ref{fig:region}, we depict the stable throughput region $\mathbb{R}_D$ of the system $\mathcal{S}_D$. The boundary between the energy-limited part and the interference limited part, from the perspective of the Type II node is $\lambda_1 = \tilde{\lambda}_1$, where $\tilde{\lambda}_1$ is the arrival rate for $Q_1$ at which both data queues, $ Q_1$ and $Q_2$, become saturated. Accordingly, the energy-limited sub-region is characterized by 
\begin{align}
\mathbb{R}_{D |  \lambda_1 \leq \tilde{\lambda}_1 }=\big\{ (\lambda_1,\lambda_2) \in \mathbb{R}_D | \ \lambda_1 \leq \tilde{\lambda}_1 \big\},
\end{align}
and the interference limited sub-region is characterized by  
\begin{align}
\mathbb{R}_{D |  \lambda_1 > \tilde{\lambda}_1 }=\big\{ (\lambda_1,\lambda_2) \in \mathbb{R}_D | \ \lambda_1 > \tilde{\lambda}_1 \big\}. 
\end{align}

%

\subsection{The Generalized Stochastic Dominance Approach}
In this section, we rely on stochastic dominance arguments \cite{rao1988stability}, which are instrumental in establishing the stable throughput region of $\mathcal{S}_D$. However, the conventional stochastic dominance approach should be modified for our system, since the transmission of dummy packets by Type I node increases the harvesting opportunities for Type II node. Thus, in order to construct a hypothetical ``dominant" system in which the queue lengths are never smaller than their counterparts in the system $\mathcal{S}_D$, the hypothetical system proposed in \cite{rao1988stability} is modified to capture the two-mode operation inherent to our RF EH system.\\ 
Recall, from (\ref{eqn: int mu2}), that the service rate of $Q_2$ increases with $\lambda_1$. Thus, it is straightforward to show that, using classic stochastic dominance arguments, saturating $Q_1$ increases the service rate of $Q_2$. Hence, the queue length in this hypothetical system (particularly $Q_2$) no longer dominates (\.i.e. could be smaller) its counterpart in the system $\mathcal{S}_D$ and, thus, the classic stochastic dominance argument fails. For example, if we consider the case where $\lambda_1 = 0$ and $\lambda_2 >0$, we observe that $\lambda_2 \leq q_2 $ belongs to the stable throughput region, which contradicts (\ref{eqn: int mu2}), where for $\lambda_1=0$, we get $\lambda_2=0$.

Hence, we define the hypothetical system for our RF EH system to be constructed as follows: 
\begin{itemize}
\item Arrivals at data queues $Q_1$ and $Q_2$ occur at the same instants as in the system $\mathcal{S}_D$.
\item Transmission decisions, determined by independent coin tosses, are identical to those in the system $\mathcal{S}_D$.
\item In the energy-limited region, Type I node does not transmit dummy packets. Thus, the energy arrivals to the battery queue of the harvesting node occurs exactly at the same instants as in the system $\mathcal{S}_D$.
\item In the interference-limited region, if the data queue is empty and the node decides to transmit, a dummy packet is transmitted, if the node has sufficient energy to transmit.  
\end{itemize}

\indent From the construction of the new hypothetical system proposed above, it can be noticed that it behaves like the system $\mathcal{S}_D$ in the energy-limited region and dominates the system $\mathcal{S}_D$ in the interference-limited region. Thus, this hypothetical system dominates our system $\mathcal{S}_D$ because the transmissions of dummy packets collide with the transmission of the other node. Also, the transmissions of dummy packets consume energy without contributing to the throughput of Type II node.
\vspace{-0.15in}
%
\subsection{Establishing the Stability Conditions of $\mathcal{S}_D$} \label{subsec: conditions}
In order to derive the stability conditions of the system $\mathcal{S}_D$, we construct two dominant systems, where in the first dominant system Type II node is backlogged and in the second dominant system, Type I node is backlogged only in the interference-limited region. 
\subsubsection{First dominant system}
\indent In this hypothetical system, we consider the case where the Type II node continues transmitting dummy packets whenever its data queue, $Q_2$, is empty given that its battery is non-empty. Since the system is interference-limited from the perspective of Type I node, our dominant system is identical to the one proposed in \cite{rao1988stability}. Hence, the saturated service rate of $Q_1$ is given by
\begin{equation}
\mu_1^s = q_1 \big( 1- \min \big\lbrace \frac{q_1 p_h}{ (1+ q_1 p_h)}, q_2 \big\rbrace \big). 
\end{equation}
Also, by substituting $\mathbb{P}[Q_1 >0 | Q_2 >0] =\lambda_1 / \mu_1^s$ in (\ref{eqn: int mu2}), the service rate of $Q_2$ becomes
\begin{equation}
\mu_2 = \dfrac{(1-q_1) \min \big\lbrace \frac{q_1 p_h}{ (1+ q_1 p_h)}, q_2 \big\rbrace \lambda_1}{q_1 \left( 1- \min \big\lbrace \frac{q_1 p_h}{ (1+ q_1 p_h)}, q_2 \big\rbrace \right)}.
\end{equation}
Therefore, the stable throughput region of the first dominant system $ \mathbb{S}_1$ is given by 
\begin{align}\label{eqn: S1}
\mathbb{R}_1 = \bigg\{ (\lambda_1, \lambda_2) \ \big| \ \lambda_1 \leq q_1 \big( 1- \min \big\lbrace \frac{q_1 p_h}{ (1+ q_1 p_h)}, q_2 \big\rbrace \big), \ \lambda_2 \leq  \dfrac{(1-q_1) \min \left\lbrace \frac{q_1 p_h}{ (1+ q_1 p_h)}, q_2 \right\rbrace \lambda_1}{q_1 \left( 1- \min \left\lbrace \frac{q_1 p_h}{ (1+ q_1 p_h)}, q_2 \right\rbrace \right)}  \bigg\}. 
\end{align}

Also, since the system becomes interference-limited from the perspective of node II, when the data queues, $Q_1$ and $Q_2$, are backlogged, $\tilde{\lambda}_1$ is given by 
\begin{equation}
\tilde{\lambda}_1 = \mu_1^s = q_1 \left( 1- \min \left\lbrace \frac{q_1 p_h}{ (1+ q_1 p_h)}, q_2 \right\rbrace \right). 
\end{equation}
\subsubsection{Second dominant system}
\indent In this hypothetical system, $Q_1$ is backlogged only in the interference-limited region from the perspective of Type II node. 
In the interference-limited part of $\mathbb{R}_{2}$, the saturated service rate of $Q_2$ is given by 
\begin{equation}
\mu_2^s= (1-q_1) \min \left\lbrace \frac{q_1 p_h}{ (1+ q_1 p_h)}, q_2 \right\rbrace.
\end{equation}
Similarly, by substituting $\mathbb{P}[Q_2 >0 | Q_1 >0] =\lambda_2 / \mu_2^s$ in (\ref{eqn: int mu1}), we obtain 
\begin{equation}\label{eqn: mu1 2nd dom}
\mu_1= q_1 \left( 1- \frac{\lambda_2}{1-q_1}\right). 
\end{equation}
Therefore, the stable throughput region of the interference-limited part of the second dominant system is given by 
\begin{align}
\mathbb{R}_{2 |  \lambda_1 \geq \tilde{\lambda}_1 } = \big\{   (\lambda_1, \lambda_2) \ | \  \tilde{\lambda}_1  \leq \lambda_1 \leq  q_1 \big( 1- \frac{\lambda_2}{1-q_1}\big), \lambda_2 \leq  (1-q_1) \min \big\lbrace \frac{q_1 p_h}{ (1+ q_1 p_h)}, q_2 \big\rbrace  \big\}.  
\end{align}
The stable throughput regions of the dominant systems $\mathbb{R}_1$ and $\mathbb{R}_{2 |  \lambda_1 \geq \tilde{\lambda}_1 }$ are shown in Fig. \ref{fig:region}. 
\subsubsection{Stability region of the system $\mathcal{S}_D$}
In the following lemma we derive the relationship between the stable throughput region of the dominant systems $\mathbb{R}_1$ and $\mathbb{R}_{2 |  \lambda_1 \geq \tilde{\lambda}_1 }$ and the original system $\mathbb{R}_D$.\\ 
\indent \textbf{Lemma 3.} The stable throughput region $\mathbb{R}_D$ of the system $\mathcal{S}_D$ is given by the union of the stable throughput region of the first dominant system and the interference-limited part of the second dominant system, \.i.e., 
$
\mathbb{R}_D= \mathbb{R}_1 \cup \mathbb{R}_{2 |  \lambda_1 \geq \tilde{\lambda}_1 }.
$
\begin{proof}
The stable throughput region of the original system is the union of the two dominant systems, based on \cite{rao1988stability}, \.i.e., $\mathbb{R}_D= \mathbb{R}_1 \cup \mathbb{R}_{2}$. From the construction of the second dominant system, we know that the energy-limited region is identical to the original system, \.i.e., $\mathbb{R}_{2 |  \lambda_1 < \tilde{\lambda}_1 } = \mathbb{R}_{D |  \lambda_1 < \tilde{\lambda}_1 }$. Hence, we have $\mathbb{R}_2 = \mathbb{R}_{D |  \lambda_1 < \tilde{\lambda}_1 } \cup \mathbb{R}_{2 |  \lambda_1 \geq \tilde{\lambda}_1 }$ and $ \mathbb{R}_D= \mathbb{R}_1 \cup \mathbb{R}_{D |  \lambda_1 < \tilde{\lambda}_1 } \cup \mathbb{R}_{2 |  \lambda_1 \geq \tilde{\lambda}_1 }.
$\\ 
\indent Now, assume that the rate pair $(x_1, x_2) \in \mathbb{R}_{D |  \lambda_1 < \tilde{\lambda}_1 }$. Hence, $ x_1 < \tilde{\lambda}_1 $ and $x_2 \leq \mu_2$. Since we achieve the maximum service rate for Type II node by backlogging $Q_2$, from (\ref{eqn: S1}) we obtain 
\begin{align}
x_2 \leq \dfrac{(1-q_1) \min \left\lbrace \frac{q_1 p_h}{ (1+ q_1 p_h)}, q_2 \right\rbrace x_1}{q_1 \left( 1- \min \left\lbrace \frac{q_1 p_h}{ (1+ q_1 p_h)}, q_2 \right\rbrace \right)}. 
\end{align}
Therefore, the rate pair $(x_1, x_2) \in \mathbb{R}_1$ and $ \mathbb{R}_{D |  \lambda_1 < \tilde{\lambda}_1 } \subseteq \mathbb{R}_1 $. 
\end{proof}
\indent \textbf{Proposition 1.} The stability region of the system $ \mathcal{S}_D$ is given by 
\begin{align}\label{eqn: prop 1 } 
\mathbb{R}_D= \bigg\{  (\lambda_1, \lambda_2) \ \big | \ \lambda_1 \leq  q_1 \left( 1- \frac{\lambda_2}{1-q_1}\right),  \lambda_2 \leq  \dfrac{(1-q_1) \min \left\lbrace \frac{q_1 p_h}{ (1+ q_1 p_h)}, q_2 \right\rbrace \lambda_1}{q_1 \left( 1- \min \left\lbrace \frac{q_1 p_h}{ (1+ q_1 p_h)}, q_2 \right\rbrace \right)} \bigg\}. 
\end{align}
\begin{proof}
For the purpose of the proof, we also define a conventional Aloha system with nodes with unlimited energy supplies, i.e., $\mathcal{S}_{\text{Aloha}} $.
The outline of the proof is as follows \begin{enumerate}[(i)] \item The generalized stochastic dominance approach proves the necessity of the stability conditions on $Q_1$ and $Q_2$ in (\ref{eqn: prop 1 }). Meanwhile, it only proves the sufficiency of the stability conditions for $Q_2$.  \item The stability condition on $ Q_1$ in $\mathcal{S}_{\text{Aloha}} $ is sufficient for stability of $Q_1$ in $\mathcal{S}_D$. \item The stability condition on $Q_1$ is the same in both systems $\mathcal{S}_{\text{Aloha}} $ and $\mathcal{S}_D $. \item From (ii) and (iii), we establish the sufficiency of the stability condition on $Q_1$ for $\mathcal{S}_D$.   \end{enumerate}

\textit{The detailed proof is as follows}
\begin{enumerate}[(i)]
\item Recall that for systems with unlimited energy such as $ \mathcal{S}_{\text{Aloha}}$, the sufficient and necessary stability conditions are given by the union of the stability regions of the two hypothetical systems in \cite{rao1988stability}. On the other hand, for a system with batteries the transmission of dummy packets in the hypothetical system wastes the energy of the nodes which limits the data transmissions in future slots. For instance, in a hypothetical system where \emph{the first node} is backlogged, there may exist instants at which \emph{the first node} is unable to transmit due to energy outage, while it is able to transmit in the original system. Hence, \emph{the second node} in the hypothetical system may have a higher success rate compared to the original system, and the sample path dominance is violated. Consequently, the union of the stability conditions of two hypothetical systems is only a necessary condition for the stability of the original system \cite{jeon2015stability}. 
In our paper, we arrived to (\ref{eqn: prop 1 }) by applying the generalized stochastic dominance approach, in which the first dominant system is constructed such that Type II node continues transmitting dummy packets whenever its data queue, $Q_2$, is empty. The transmission of dummy packets from Type II node in this system wastes energy. Therefore, Type I node in this dominant system may have a better success rate compared to that of the system $\mathcal{S}_D$. Therefore, the generalized stochastic dominance approach only proves the necessity of the stability condition on $Q_1$ in (\ref{eqn: prop 1 }). On the other hand, the stability condition on $Q_2$ is sufficient and necessary, since Type I node has unlimited energy and the previous argument only applies to nodes with batteries.

\item In $\mathcal{S}_{\text{Aloha}} $ the two contending nodes have unlimited energy, while in $ \mathcal{S}_D$ the transmissions of the Type II node are constrained by the energy in the battery. Hence, the second contending node in $\mathcal{S}_{\text{Aloha}} $, transmits more frequently than a Type II node with the same transmission probability $q_2$ in $ \mathcal{S}_D$. Consequently, there are more collisions in $\mathcal{S}_{\text{Aloha}} $ compared with those in $\mathcal{S}_D $ and the service rate of Type I node in $ \mathcal{S}_D$ cannot be smaller than the service rate of the first contending node in $\mathcal{S}_{\text{Aloha}} $. Therefore, the stability condition on $ Q_1$ in $\mathcal{S}_{\text{Aloha}} $ is sufficient for stability of $Q_1$ in $\mathcal{S}_D$.

\item In the stability analysis of $\mathcal{S}_D$, we proved that the stability condition on $Q_1$ is $\lambda_1 \leq  q_1 \left( 1- \frac{\lambda_2}{1-q_1}\right)$, which is identical to the stability condition on $Q_1$ in $\mathcal{S}_{\text{Aloha}} $ \cite{rao1988stability}.

\item The sufficiency of the condition $\lambda_1 \leq  q_1 \left( 1- \frac{\lambda_2}{1-q_1}\right)$ for the stability of $ Q_1$ in $\mathcal{S}_D$ follows from its sufficiency for $ Q_1$ in $\mathcal{S}_{\text{Aloha}} $. Therefore, the stability conditions in (\ref{eqn: prop 1 }) are necessary and sufficient conditions and the region $\mathbb{R}_D $ is the exact stability region of the system $ \mathcal{S}_D$.
\end{enumerate}
\end{proof}
\vspace{-0.3in}
%
\subsection{The Relationship between Stability Regions of $\mathcal{S}_G$ and $\mathcal{S}_D$}\label{subsec: rel}
\indent \textbf{Lemma 4.} The stability region of the system $\mathcal{S}_D$ is an inner bound on the stability region of the system $\mathcal{S}_G$, i.e., $\mathbb{R}_G \supseteq \mathbb{R}_D$.  
\begin{proof}
According to the assumption $\mathbb{P}[B\!>\!0, Q_1\!=\! 0| Q_2\!>\!0]\!=\!0$ in the system $\mathcal{S}_D $, a Type II node has a lower service rate compared to a Type II node in the system $\mathcal{S}_G $. Hence, the length of $Q_2$ in the system $\mathcal{S}_D$ is never smaller than its counterpart in the system $\mathcal{S}_G$, i.e., the system $\mathcal{S}_D$ dominates $\mathcal{S}_G$ from the perspective of Type II node. Consequently, the stability of $Q_2$ in $\mathcal{S}_D $, is sufficient for the stability of $Q_2$ in $\mathcal{S}_G$. Additionally, the assumption $\mathbb{P}[B\!>\!0, Q_1\!=\! 0| Q_2\!>\!0]\!=\!0$, implies that Type II node is not transmitting in the time slots at which $ Q_1$ is empty. Hence, the number of idle slots increases, since at those time slots Type I node has no data packets to transmit. Accordingly, the service rate of Type I node is not affected, and the stability condition on $Q_1$ is the same in $\mathcal{S}_D$ and $\mathcal{S}_G$. We conclude that the stability region of $\mathcal{S}_D$ is an inner bound on the stability region of $\mathcal{S}_G$.
\end{proof}
Finally, from Proposition 1 and Lemma 4 we arrive to Theorem 1. 

%
\subsection{The Closure over all Transmission Probabilities}\label{subsec: closure}
\begin{figure}[t]
\includegraphics[width=0.55\textwidth, height=0.25\textheight]{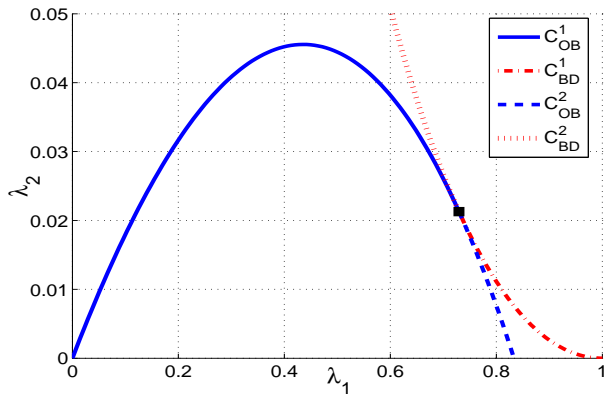}
\centering
\caption{The closure $\mathcal{C}_{(\mathbb{R}_G^{\text{inner}})}= \mathcal{C}_{\textit{OB}}^1 \cup \mathcal{C}_{\textit{BD}}^1$, for $p_h=0.2$. }\label{fig: closure}
\end{figure}

In this subsection, we prove Theorem 2. The closure of the inner bound $\mathbb{R}_G^{\text{inner}}$, is defined by the union of all stability regions for a given $(q_1,q_2)$, \.i.e., 
$
\mathcal{C}_{(\mathbb{R}_G^{\text{inner}})} = \bigcup\limits_{(q_1, q_2) } \ \mathbb{R}_G^{\text{inner}} ( (q_1,q_2)). 
$
In $ \mathbb{R}_G^{\text{inner}}$, the service rate of Type II node is always lower than the arrival rate of Type I node, \.i.e., $\mu_2 < \lambda_1$. Also, from (\ref{eqn thm1}), we note that $\mu_2$ is increasing in $\min \lbrace \frac{q_1 p_h}{ (1+ q_1 p_h)}, q_2 \rbrace $, while $\mu_1$ is not affected by $q_2$. Hence, in order to find the closure $\mathcal{C}_{(\mathbb{R}_G^{\text{inner}})}$, we need to find $q_2$ that maximizes $\mu_2$. From (\ref{eqn thm1}), we observe that for maximizing $\mu_2$, the transmission probability of Type II node $q_2$ should be greater than or equal to $q_2^* = (q_1 p_h)/(1+q_1 p_h)$. Also, increasing $q_2$ beyond $q_2^*$ does not affect the service rate $\mu_2$.\\
\indent Interestingly, we can interpret $q_2^*$ using \emph{Renewal reward theorem} \cite{ross1996stochastic}. Assume that the data queues are backlogged and Type II node transmits whenever it receives an energy packet, \.i.e., $q_2=1$. Let the expected reward $R$ that Type II node obtains, be the transmission of one data packet, \.i.e., $\mathbb{E}[R]=1$. Also, the expected number of time slots, $T$, needed for the transmission of one data packet is one slot for transmission, and $(q_1 p_h)^{-1}$ slots are needed for harvesting one energy packet. Hence, the expected time needed for a transmission $\mathbb{E}[T]=1 + (q_1 p_h)^{-1}$. Using the renewal reward theorem, we find that the effective transmission rate of Type II node is given by 
$
q_{2, \textit{eff}}= \frac{\mathbb{E}[R]}{\mathbb{E}[T]}=\frac{q_1 p_h}{1+q_1 p_h}.
$
Therefore, the previous expression represents the maximum possible transmission rate of Type II node, which is the minimum  transmission probability $q_2$ that maximizes $\mu_2$. Now, the problem of finding the closure $\mathcal{C}_{(\mathbb{S}_o)}$, reduces to finding the closure of $ \mathbb{R}_G^{\text{inner}} ( (q_1,q_2^{*}) )$ over all $q_1$, \.i.e., 
$
\mathcal{C}_{(\mathbb{R}_G^{\text{inner}})} = \bigcup\limits_{q_1 \in [0,1]} \ \mathbb{R}_G^{\text{inner}} ( (q_1,q_2^{*})), 
$
where 
\begin{align}
\mathbb{R}_G^{\text{inner}}( (q_1,q_2^{*})) =\big\{  (\lambda_1, \lambda_2) \big| \ \lambda_1 \leq  q_1 \big( 1- \frac{\lambda_2}{1-q_1}\big), \lambda_2 \leq (1-q_1) p_h \lambda_1 \big\}, 
\end{align}
which represents the triangle OBD in Fig. \ref{fig: region finite}, where $ \textit{D}\!=\!(q_1,0)$, and $\textit{B}\!=\!\left( \frac{q_1}{1+q_1 p_h} ,\frac{q_1 p_h (1-q_1)}{1+q_1 p_h} \right) \!.$
\indent Since, we know that the region $\mathbb{R}_G^{\text{inner}}$ consist of two line segments, the closure $\mathcal{C}_{(\mathbb{R}_G^{\text{inner}})}$ can be found be taking the union of the closures of the line segments $\overline{\textit{OB}}$, $\overline{\textit{BD}}$. First, we find the closure of the line segment $\overline{\textit{OB}}$ by solving $x= \frac{q_1}{1+q_1 p_h}$, and $y=\frac{q_1 p_h (1-q_1)}{1+q_1 p_h}$. The solution represents the trace of the point \textit{B} for $q_1 \in [0,1]$, which is $y=p_h x (1- \frac{x}{1-p_h x})$. Hence, the closure of $\overline{\textit{OB}}$ is given by 
\vspace{-0.3in}
\begin{align}
\mathcal{C}_{\overline{\textit{OB}}}=\big\{  (\lambda_1, \lambda_2) \big| \ \lambda_2 \leq  \lambda_1 p_h \big( 1- \frac{\lambda_1}{1-\lambda_1 p_h}\big) \big\},
\end{align} 
which is represented in Fig. \ref{fig: closure} by $\mathcal{C}_{\textit{OB}}^1$ and $\mathcal{C}_{\textit{OB}}^2$. Note that $\mathcal{C}_{\overline{\textit{OB}}}$ is a convex region, since it is a hypograph of a concave function. 
 Next, in order to find the closure of $\overline{\textit{BD}}$, we solve $ \max\limits_{q_1 \in [0,1]} \ \ q_1 \big(  1-\frac{\lambda_2}{1-q_1}\big).  $ The solution gives us the closure of the line segment extending $\overline{\textit{OB}}$ to the $\lambda_2$-axis, represented by $\mathcal{C}_{\textit{BD}}^1$ and $\mathcal{C}_{\textit{BD}}^2$ in Fig. \ref{fig: closure}. However, since we want the closure of $\overline{\textit{BD}}$ and not the extension, $ \mathcal{C}_{\overline{\textit{BD}}}$ is limited from the left by the trace of the point $\textit{B}$. Thus, the closure of $\overline{\textit{BD}}$ is the region bounded from the left by $\mathcal{C}_{\textit{OB}}^2$ and bounded by $\mathcal{C}_{\textit{BD}}^1$ from the right. It is characterized by 
\begin{align} 
&\mathcal{C}_{\overline{\textit{BD}}}=\big\{  (\lambda_1, \lambda_2) \ \big| \ \sqrt{\lambda_1}+\sqrt{\lambda_2} \leq 1, \ \lambda_1 \! > \! \dfrac{1 \! + \! 2 p_h \! - \! \sqrt{1 \! + \!4 p_h}}{2 p_h^2},\ \lambda_2 \! > \! \lambda_1 p_h   \big( 1 \! - \! \frac{\lambda_1}{1 \! - \! \lambda_1 p_h}\big)   \big\} . 
\end{align} 
Note that the second condition can be found by solving the two equations of $\mathcal{C}_{\textit{OB}}^2$ and $\mathcal{C}_{\textit{BD}}^1$. Finally, the closure $\mathcal{C}_{(\mathbb{R}_G^{\text{inner}})}$ is the union of $\mathcal{C}_{\textit{OB}}$ and $\mathcal{C}_{\textit{BD}}$, which is represented by $\mathcal{C}_{\textit{OB}}^1$ and $\mathcal{C}_{\textit{BD}}^1$ in Fig. \ref{fig: closure}. After some algebraic manipulations, we obtain (\ref{eqn: closure}).\\
\indent \textbf{Remark:} (\ref{eqn: closure}) can also be obtained by maximizing the service rate of Type II node $\mu_2$, subject to the stability condition of Type I node $\lambda_1 \leq \mu_1$, \.i.e, 
\begin{align}
\max\limits_{q_1 \in [0,1]} \  (1-q_1) p_h \lambda_1  \ \ \textit{s.t } \ \lambda_1 \leq  q_1 \big( 1- \frac{\lambda_2}{1-q_1}\big).
\end{align}
Hence, the closure of our system is equivalent to maximizing $\mu_2$, because $\mu_2 $ is upper bounded by $\mu_1$. Thus, by maximizing $\mu_2$, we implicitly maximize $\mu_1$.  
%
%
\section{Model Extensions} \label{sec:Extensions}
In this section, we discuss two extensions to the previous stability analysis. First, we investigate the impact of having a finite capacity battery on the stable throughput region. Second, we investigate the effect of having a full-duplex RF energy harvesting node.

\subsection{Finite Capacity Battery}
\begin{figure}[t]
\includegraphics[width=0.4\textwidth, height=0.2\textheight]{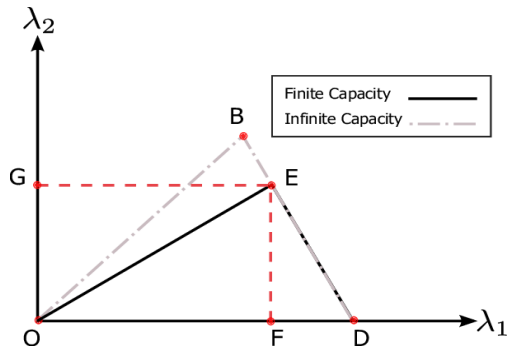}
\centering
\caption{The stable throughput region of opportunistic energy harvesting Aloha with finite capacity battery is characterized by the triangle OED.}\label{fig: region finite}
\end{figure}

In this subsection, we investigate the impact of having a finite capacity battery on the stable throughput region obtained in Theorem 1. Let $M$ be the capacity of Type II node battery. Thus, the battery evolution equation becomes 
\begin{equation}
B^{t+1}=\min \left\lbrace B^{t}-\mu_B^{t}+H^{t}, M \right\rbrace 
\end{equation}
Similar to the unlimited battery capacity case, the battery queue forms a decoupled discrete time Markov chain given that the data queues are backlogged. By analyzing the Markov chain we find the probability that the battery queue is non-empty.\\
\indent \textbf{Lemma 5.} For a half-duplex RF energy harvesting node with battery of size $M$, the probability that the battery is non-empty, $\zeta$, given the data queues $Q_1$ and $Q_2$ are backlogged, is given by 
\begin{equation}
\zeta= \begin{cases}
\rho \left( \frac{1-\left( \frac{q_1 p_h (1-q_2)}{q_2}\right)^M}{1- \rho \left( \frac{q_1 p_h (1-q_2)}{q_2}\right)^M}\right) , \ q_2 \neq \frac{q_1 p_h }{1+ q_1 p_h},\\
1,\ q_2 = \frac{q_1 p_h }{1+ q_1 p_h}
\end{cases}
\end{equation}
where $\rho=\frac{q_1 p_h}{q_2(1+q_1 p_h)}$ is the probability that the battery is non-empty in the infinite capacity battery case.
\begin{proof}
Along the lines of Lemma 2.
\end{proof}
\indent Next, we apply the same procedure used in proving the stability region for the infinite battery capacity case. \\
\indent \textbf{Corollary 1.} An inner bound on the stable throughput region of the opportunistic RF energy harvesting slotted Aloha is the triangle OED, shown in Fig .\ref{fig: region finite}. Assuming half-duplex energy harvesting and a battery of size $M$, the region is characterized by the following equation  
\begin{align} \label{eqn thm2}
\mathbb{R}_G^{\text{inner}}= \left\lbrace  (\lambda_1, \lambda_2) | \ \lambda_1 \leq  q_1 \left( 1- \frac{\lambda_2}{1-q_1}\right),  \lambda_2 \leq  \dfrac{q_2 (1-q_1) \zeta \lambda_1}{q_1 \left( 1- q_2 \zeta \right)} \right\rbrace 
\end{align}
\begin{proof}
Along the lines of Theorem 1.
\end{proof}
\indent The reduction in the stability region due to finite capacity battery is the triangle OEB shown in Fig. \ref{fig: region finite}.

\subsection{Full-Duplex Energy Harvesting}
Now, we investigate the effect of having a full-duplex RF energy harvesting Type II node, \.i.e., the harvesting circuit is separated form the transmission circuit. Hence, a node can transmit and harvest simultaneously. Also, full-duplex energy transmission is advantageous because harvesting self-interference may introduce high energy yield. In the full-duplex RF energy harvesting paradigm, we have three types of harvesting opportunities. First, harvesting the transmissions of Type I node while Type II node is silent. Similar to the half-duplex case, we model this case by a Bernoulli process with mean $p_{h|\lbrace 1 \rbrace}$. Second, harvesting the self-interference of Type II node while Type I node is silent, which is modeled by a Bernoulli process with mean $p_{h|\lbrace 2 \rbrace}$. Third,  harvesting both transmissions of Type I node and the self-interference of Type II node, which is modeled by a Bernoulli process with mean $p_{h|\lbrace1, 2 \rbrace}$.\\
\indent In order to characterize the probabilities $p_{h|\lbrace 2 \rbrace}$ and $p_{h|\lbrace1, 2 \rbrace}$, we use a similar approach to the one used in characterizing $p_{h|\lbrace 1 \rbrace}$ in Section \ref{sec:Energy_Harvesting_Model}. We assume that the loopback interference coefficient $c \in [0,1]$ is known \cite{thangaraj2012self}. Also, we assume a Rayleigh fading channel between the transmit antenna and the harvesting antenna, \.i.e., $g_t \sim \exp(1)$. In case only self-interference is present, the received power at time slot $t$ is equal to $P_{R}(t)= \eta  P_2  c g_t$. Hence, using the same approach as in the half-duplex case, we obtain
\begin{equation}
p_{h|\lbrace 2 \rbrace}= \frac{1}{1+ \frac{\gamma}{\eta P_2 c}}.
\end{equation}
In case both transmissions of Type I node and the self-interference are present, the received power at time slot $t$ is equal to $ P_{R}(t)=\eta (P_1 h_t (1+l^\alpha)^{-1} +  \ P_2 \ c \ g_t )$. The probability $p_{h|\lbrace1, 2 \rbrace}$ can be characterized in a similar fashion\footnote{In order to characterize $p_{h|\lbrace1, 2 \rbrace}$, we need the distribution of the sum of independent gamma distributed random variables, all with integer shape parameters and different rate parameters, which is called the generalized integer gamma distribution (GIG) \cite{coelho1998generalized}.}.\\
\indent For a full-duplex harvesting node, the average energy harvesting process of the battery queue is given by 
\begin{align}
\mathbb{E}[H^{t}] &= q_1 p_{h|\lbrace 1\rbrace} \mathbb{P}[Q_1>0] +  q_2 p_{h|\lbrace 2 \rbrace} \mathbb{P}[Q_1=0, B>0, Q_2>0]  \\ &+ (q_2 p_{h|\lbrace 2\rbrace} +q_1 q_2 (p_{h|\lbrace 1,2 \rbrace}-p_{h|\lbrace 2 \rbrace}-p_{h|\lbrace 1 \rbrace}) ) \mathbb{P}[Q_1>0, B>0, Q_2 >0] \nonumber, 
\end{align}
where $p_{h|\mathcal{M}}, $ is the harvesting probability given a set $\mathcal{M} $ of nodes are transmitting. The battery queue forms a decoupled Markov chain given that the data queues are backlogged. By analyzing the Markov chain we find the probability that the battery is non-empty, which is given by the following lemma.\\
\indent \textbf{Lemma 6.} For a full-duplex RF energy harvesting node with infinite capacity battery, the probability that the battery is non-empty $\Psi$, given the data queues $Q_1$ and $Q_2$ are backlogged, is given by 
\begin{equation}
\Psi= \min \left\lbrace \frac{q_1 p_{h|\lbrace 1\rbrace}}{q_2 (1-q_1 (p_{h|\lbrace 1,2 \rbrace}-p_{h|\lbrace 1\rbrace})-(1-q_1)p_{h|\lbrace 2 \rbrace} )}, 1\right\rbrace. 
\end{equation} 
\begin{proof}
Along the lines of Lemma 2.
\end{proof}
\indent We notice that the probability of non-empty battery for the full-duplex case is higher than that of the half-duplex case, \.i.e, $\Psi \geq \frac{q_1 p_{h|\lbrace 1\rbrace}}{ 1+ q_1 p_{h|\lbrace 1\rbrace}}.$ 
The stable throughput region of the system is given by\\
\indent \textbf{Corollary 2.} An inner bound on the stable throughput region of the opportunistic RF energy harvesting slotted Aloha, under full-duplex energy harvesting mode and infinite capacity battery, is the region characterized by  
\begin{align} \label{eqn thm3}
\mathbb{R}_G^{\text{inner}}= \left\lbrace  (\lambda_1, \lambda_2) | \ \lambda_1 \leq  q_1 \left( 1- \frac{\lambda_2}{1-q_1}\right),  \lambda_2 \leq  \dfrac{q_2 (1-q_1) \Psi \lambda_1}{q_1 \left( 1- q_2 \Psi \right)} \right\rbrace 
\end{align}
\begin{proof}
Along the lines of Theorem 1.
\end{proof}

%
%

\section{Numerical Analysis}\label{sec: Numerical}
\begin{figure}\label{fig: compare}
\centering
\begin{tabular}{ccc}
\subfloat[$q_2 >  \Psi$.]{
        \label{fig: compare1}
        \includegraphics[width=0.3\textwidth, height=0.22\textheight]{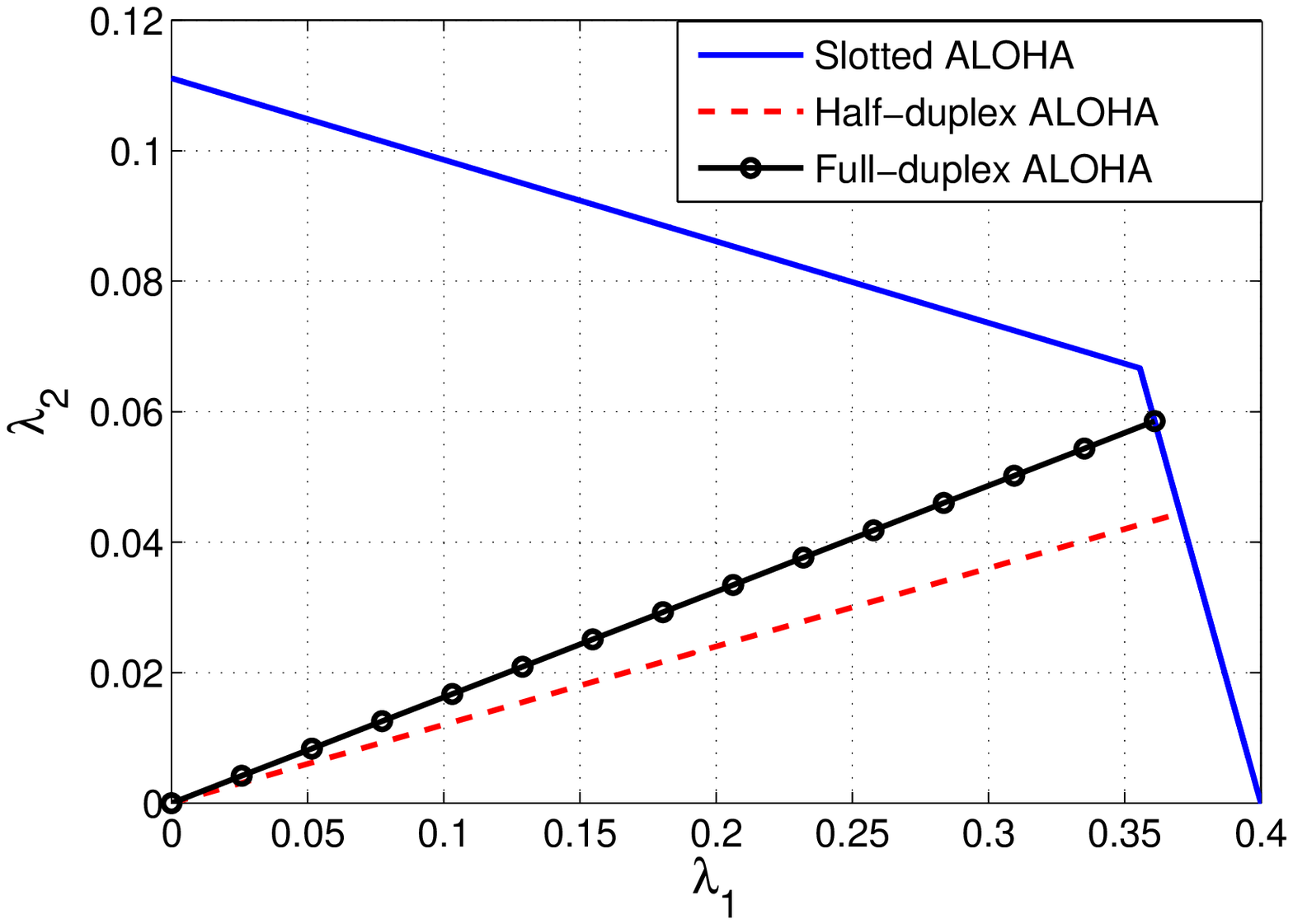} }
&
\subfloat[$ \frac{q_1 p_{h|\lbrace 1\rbrace}}{ 1+ q_1 p_{h|\lbrace 1\rbrace}} < q_2 <  \Psi$.]{
        \label{fig: compare2}
        \includegraphics[width=0.3\textwidth, height=0.22\textheight]{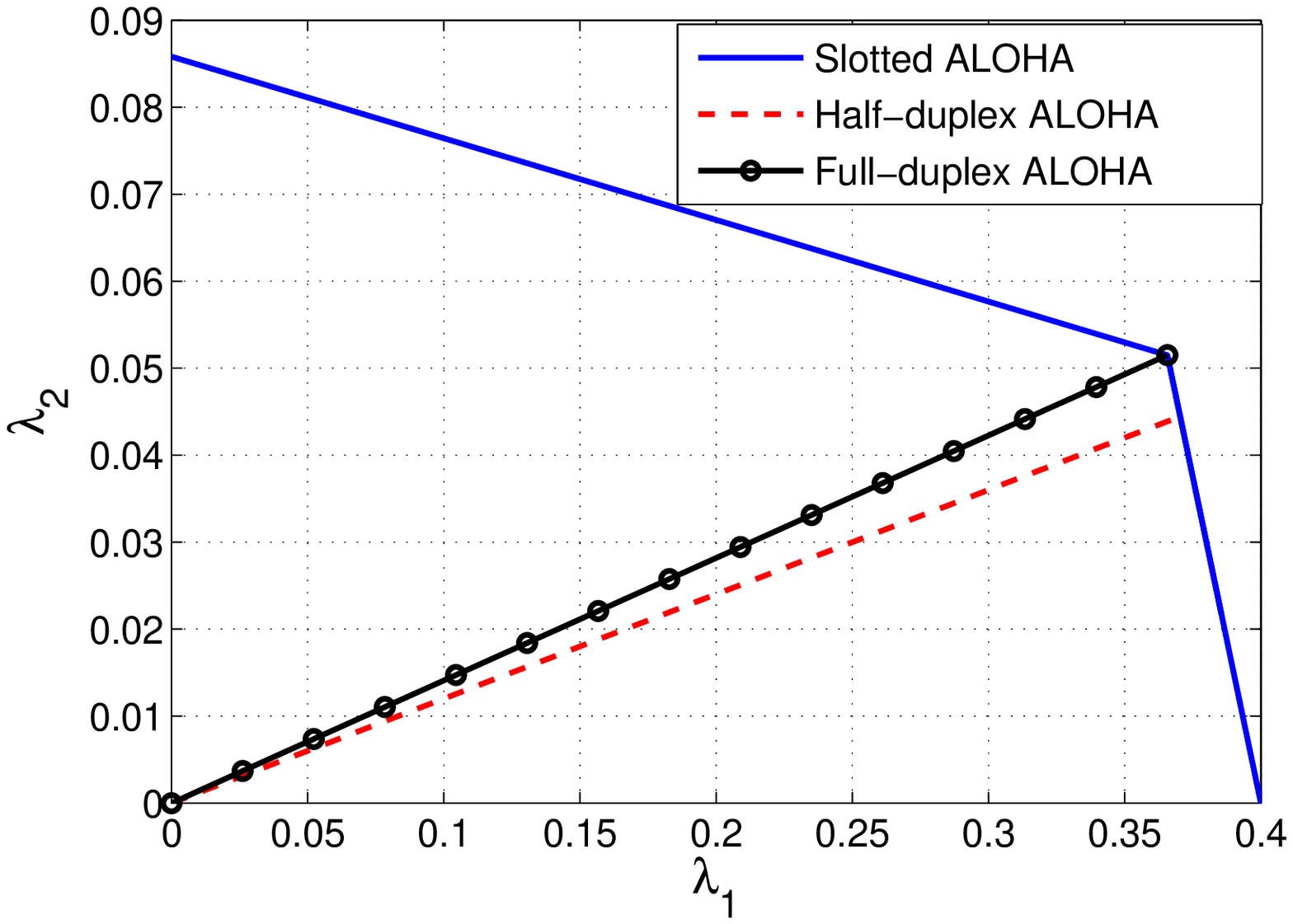} }
&
\subfloat[$q_2 < \frac{q_1 p_{h|\lbrace 1\rbrace}}{ 1+ q_1 p_{h|\lbrace 1\rbrace}}$.]{
        \label{fig: compare3}
        \includegraphics[width=0.3\textwidth, height=0.22\textheight]{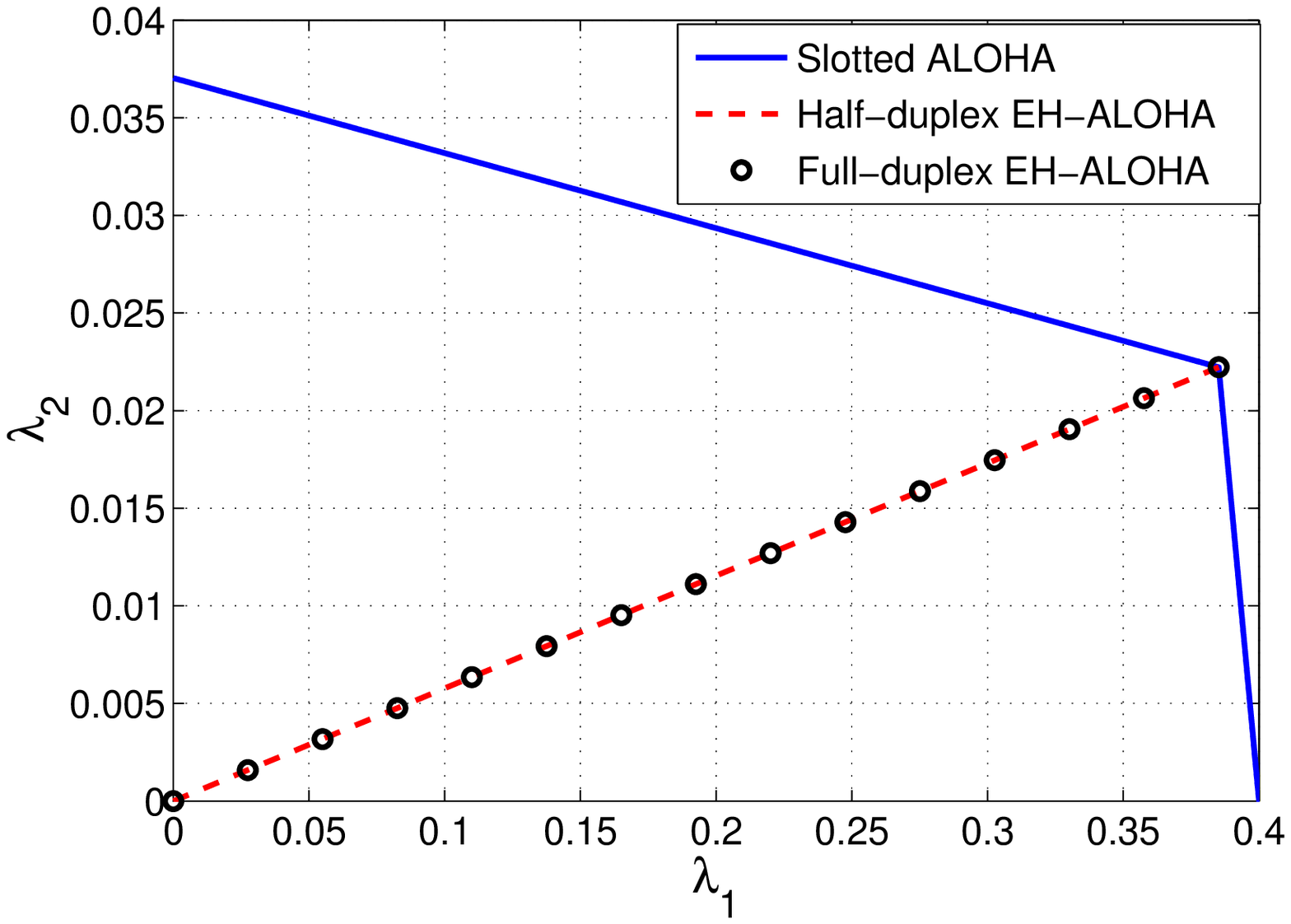} }
\end{tabular}       
\caption{The stability regions of RF EH-Aloha under half/full-duplex modes vs. slotted Aloha with unlimited energy supply.}
\label{fig: compare}
\end{figure} 
\subsection{RF EH-Aloha System $\mathcal{S}_G$ Vs. Slotted Aloha}
\indent In Fig. \ref{fig: compare}, we compare the stable throughput region of the conventional slotted Aloha with unlimited energy supply with our RF EH-Aloha system $\mathcal{S}_G$. The stability regions are shown for $q_1=0.4$, $p_{h|\lbrace 1\rbrace}=0.2$, $p_{h|\lbrace 2\rbrace}=0.2$, and $p_{h|\lbrace 1,2\rbrace}=0.35$. Also, we consider different values for $q_2$ to compare between full-duplex and half-duplex energy harvesting. We observe that the stability region of slotted Aloha is significantly reduced when RF energy harvesting is implemented, due to the energy limited sub-region. Also, for small $q_2$, \.i.e., $q_2 < \frac{q_1 p_{h|\lbrace 1\rbrace}}{ 1+ q_1 p_{h|\lbrace 1\rbrace}}$, we observe that the stability regions of half-duplex and full-duplex EH-Aloha are identical, because the service rate of Type II node is limited by the transmission probability $q_2$. 
On the other hand, for large $q_2$, \.i.e., $q_2 >  \Psi$, full-duplex RF energy harvesting expands the stability region, which agrees with intuition. From Fig. \ref{fig: compare}(a), we notice that increasing $q_2$ beyond $\Psi $ enhances the throughput of node $2$ in the slotted Aloha system. However, the throughput of the Type II node in full-duplex EH-Aloha is limited by $\Psi $.

\subsection{Validating the Stability Conditions using Simulations} \label{subsec: valid}
\begin{figure}[t]
\includegraphics[width=0.99\textwidth, height=0.3\textheight]{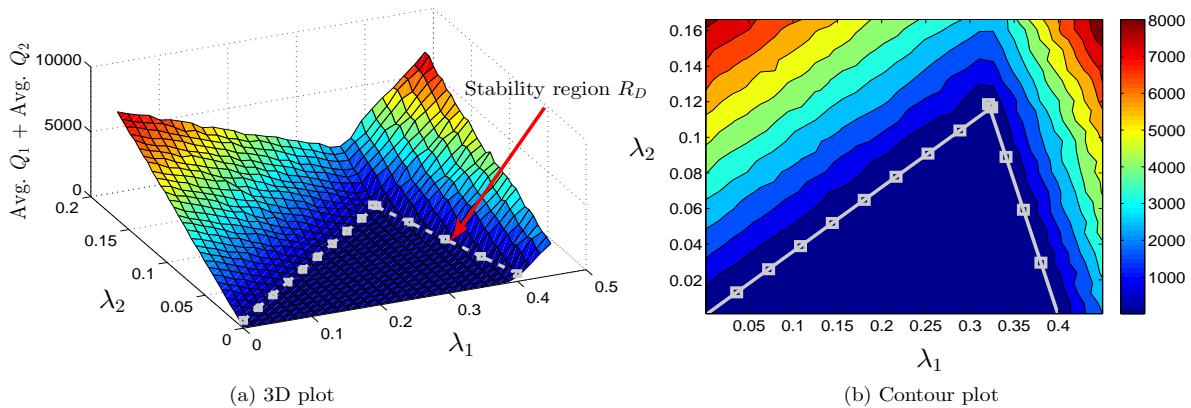} 
\centering
\caption{The stability regions $\mathbb{R}_D$ illustrated with the simulated sum average queue lenghts. }\label{fig: eval avg_Qs D}
\end{figure}

\begin{figure}[h]
\includegraphics[width=0.99\textwidth, height=0.3\textheight]{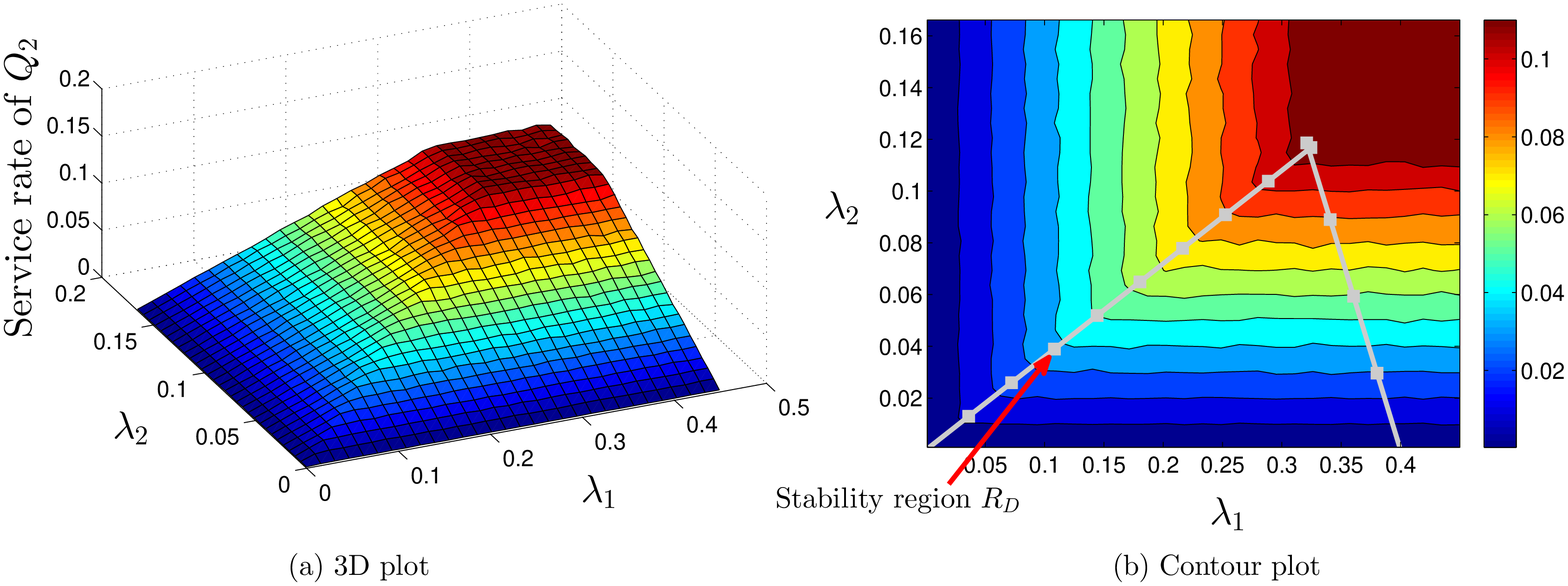} 
\centering
\caption{Simulation of the average service rate of $Q_2$ in system $\mathcal{S}_D$.}\label{fig: eval mu2 D}
\end{figure}

\begin{figure}[t]
\includegraphics[width=0.99\textwidth, height=0.3\textheight]{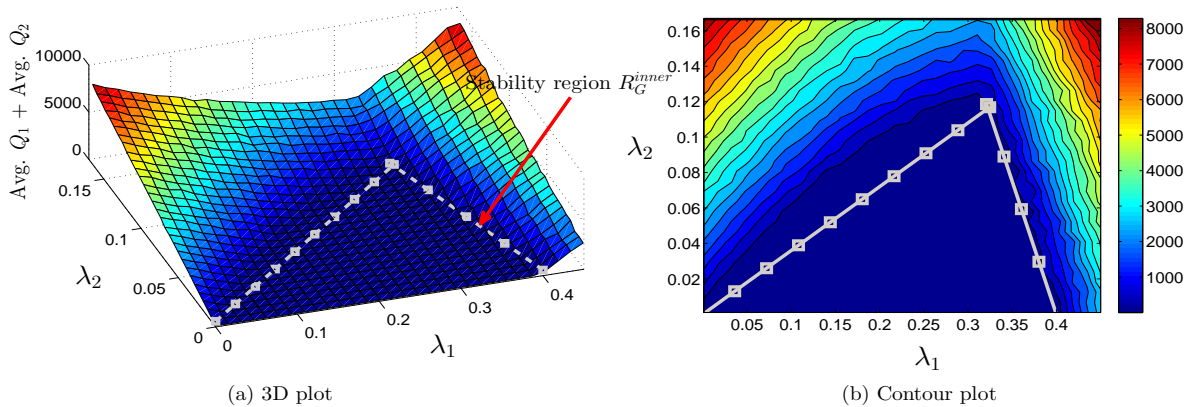} 
\centering
\caption{The stability region $\mathbb{R}_G^{\text{inner}}$ compared with the simulated sum average queue lenghts of the exact system $\mathcal{S}_O$. }\label{fig: eval avg_Qs O}
\end{figure}

\begin{figure}[t]
\includegraphics[width=0.99\textwidth, height=0.3\textheight]{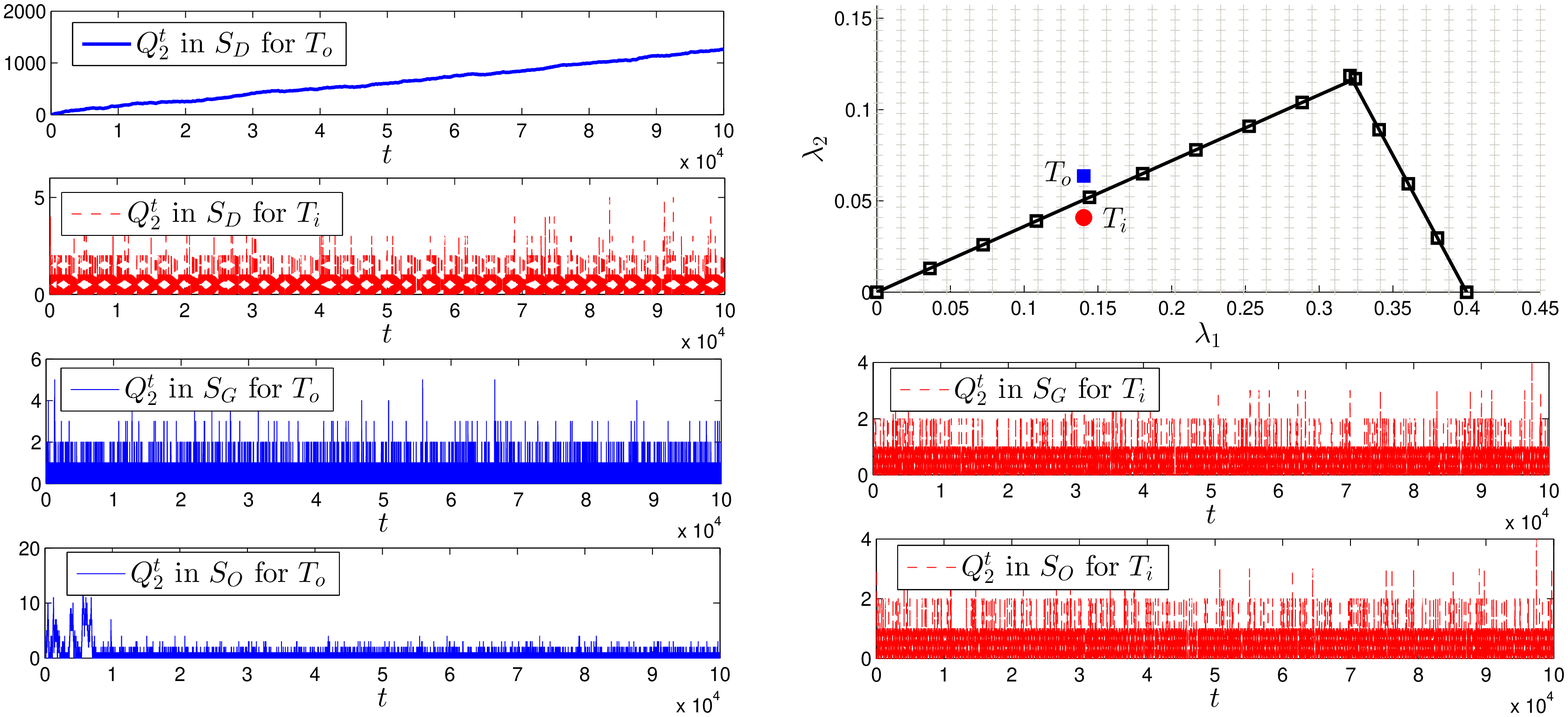} 
\centering
\caption{Sample paths for the evolution of $Q_2$ in the systems $ \mathcal{S}_D$, $ \mathcal{S}_G$ and $ \mathcal{S}_O$. }\label{fig: sample path Q2}
\end{figure}

\begin{figure}[t]
\includegraphics[width=0.99\textwidth, height=0.3\textheight]{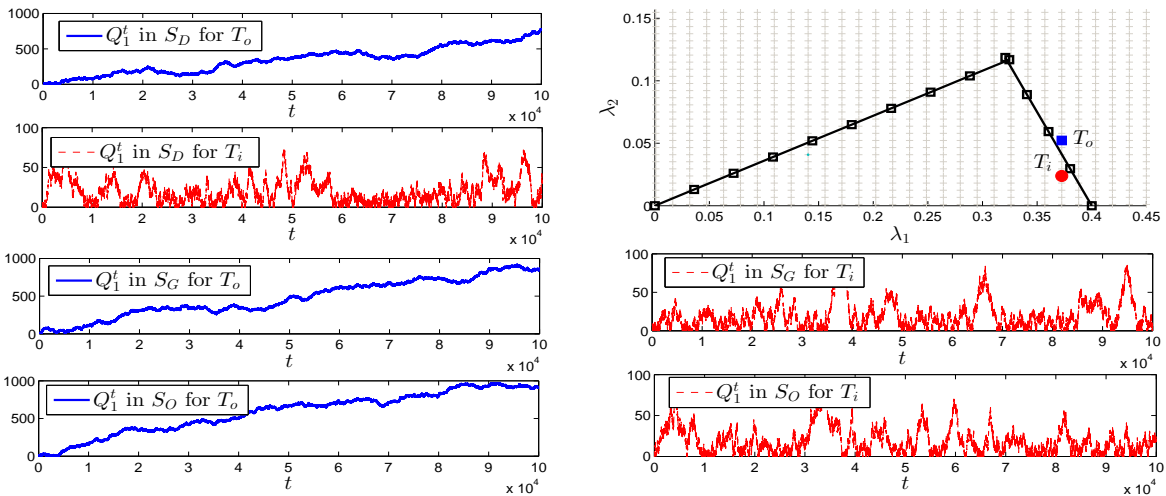} 
\centering
\caption{Sample paths for the evolution of $Q_1$ in the systems $ \mathcal{S}_D$, $ \mathcal{S}_G$ and $ \mathcal{S}_O$. }\label{fig: sample path Q1}
\end{figure}

In Fig. \ref{fig: eval avg_Qs D} and \ref{fig: eval mu2 D}, we simulate the energy harvesting Aloha system $ \mathcal{S}_D$ for $q_1\!=\!0.4$, $q_2\!=\!0.4 $, $p_h\!=\!0.6 $ and the queues are averaged over $10^5$ time slots. In particular, Fig. \ref{fig: eval avg_Qs D} shows the sum of the average queue lengths $ \mathbb{E}[Q_1^t] \!+\! \mathbb{E}[Q_2^t]$ versus the arrival rates $(\lambda_1,\lambda_2)$. We observed that the system exhibits unstable behavior, shown by the increase in the average queue lengths, as we cross the boundaries of the stability region. Fig. \ref{fig: eval mu2 D} shows the  average service rate of $Q_2$ versus the arrival rates $(\lambda_1,\lambda_2)$. It shows that the maximum service rate (lower left corner points in the contour plot), for a given $\lambda_2$, is achieved on the boundary of the stability region. These observations support that $\mathbb{R}_D$ in Proposition 1 is indeed the stability region of $ \mathcal{S}_D$.\\ 
\begin{figure}[t]
\includegraphics[width=0.99\textwidth, height=0.3\textheight]{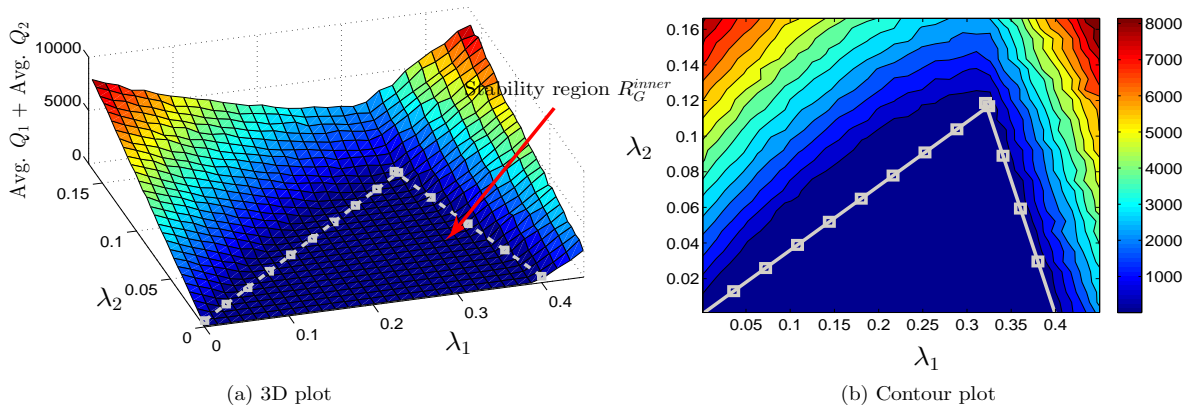} 
\centering
\caption{The stability region $\mathbb{R}_G^{\text{inner}}$ illustrated with the simulated sum average queue lenghts. }\label{fig: eval avg_Qs G}
\end{figure}
\begin{figure}
\includegraphics[width=0.99\textwidth, height=0.3\textheight]{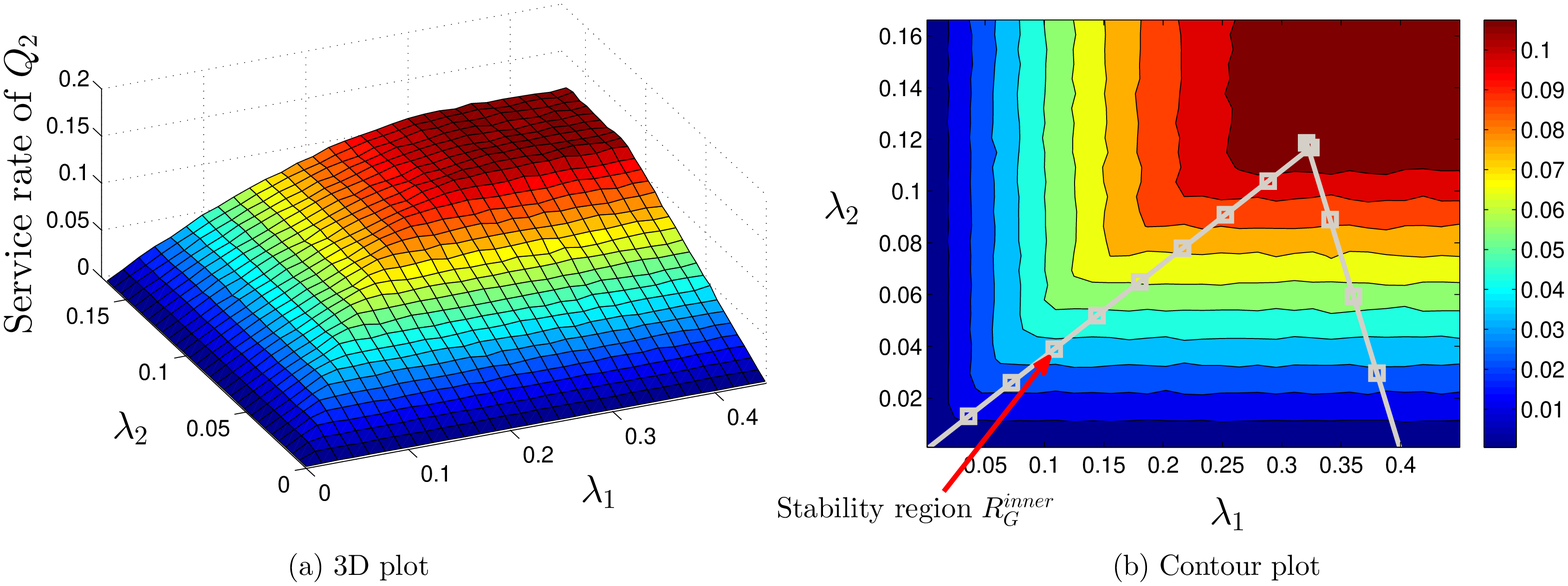} 
\centering
\caption{Simulation of the average service rate of $Q_2$ in system $\mathcal{S}_G$.}\label{fig: eval mu2 G}
\end{figure}
\indent For the aforementioned parameters, we simulate the system $ \mathcal{S}_G$ in order to verify the inner bound in Theorem 1. Similarly, the sum of the average queue lengths and the average service rate of $Q_2$ are shown in Fig. \ref{fig: eval avg_Qs G} and \ref{fig: eval mu2 G}, respectively. We notice from Fig. \ref{fig: eval avg_Qs G} that there exist rate pairs outside the left hand side of the stability region for which the queues exhibit a stable behavior. Additionally, Fig. \ref{fig: eval mu2 G} suggests that the maximum service rate (lower left corner points in the contour plot), for a given $\lambda_2$, is achieved outside the stability region. These observations indicates that $\mathbb{R}_G^{\text{inner}}$ is an inner bound on the stability region of $ \mathcal{S}_G$, as proposed in Theorem 1. \\
\indent Finally, we show the utility of the stability region $\mathbb{R}_G^{\text{inner}}$, for the exact system $\mathcal{S}_O$ which is described in Section \ref{sec: exact system}. We simulate the exact behavior of the system for $ \eta\!=\!0.7, \ \gamma\!=\!0.2335$, $P_1\!=\!1 $ and $ (1\!+\!l^\alpha)^{-1}\!=\!0.5$. Hence, $ \theta\!=\!0.667$ and from (\ref{eqn: theta}) we get $ p_h\!=\!0.6 $. The sum average queue lengths is shown in Fig. \ref{fig: eval avg_Qs O} versus the arrival rates $(\lambda_1,\lambda_2)$. We observe that the behavior of the average queue lengths in $ \mathcal{S}_O$ is very similar to $ \mathcal{S}_G$. Henceforth, we conjecture that the stability region $\mathbb{R}_G^{\text{inner}}$ is also an inner bound for the stability region of the exact system $\mathbb{R}_O$. To further support our analytical results, Fig. \ref{fig: sample path Q2} and \ref{fig: sample path Q1} present sample paths for the evolution of the queues $Q_1$ and $Q_2$ in the systems $\mathcal{S}_D$, $\mathcal{S}_G$ and $\mathcal{S}_O$. Note that the sample path of the evolution of an unstable queue should show an increasing tendency such that the queue size grow unboundedly as time increases. In Fig. \ref{fig: sample path Q2}, we show the evolution of $Q_2$ for two arrival rate pairs $T_o $ and $T_i$. We observe that $Q_2$ exhibits unstable behavior at $T_o$ in $\mathcal{S}_D$, while it exhibits stable behavior in $\mathcal{S}_G$ and $\mathcal{S}_O$. This supports our claim that the stability condition on $Q_2$ in (\ref{eqn thm1}) is sufficient and necessary in $\mathcal{S}_D$, while it is only sufficient in $\mathcal{S}_G$ and $\mathcal{S}_O$. While, Fig. \ref{fig: sample path Q1} suggests that the stability condition on $Q_1$ in (\ref{eqn thm1}) is sufficient and necessary in $\mathcal{S}_D$, $\mathcal{S}_G$ and $\mathcal{S}_O$.

\section{Conclusion} \label{sec:conclusions}
In this paper, we studied the effects of opportunistic RF energy harvesting on the stability of a slotted Aloha system consisting of a Type I node, which has unlimited energy supply, and a Type II node, which is solely powered by an RF energy harvesting circuit. We illustrated the intricacy in analyzing the exact behavior of such systems and proposed an equivalent system for which we were able to derive analytical results. In particular, we characterized an inner bound on the stability region under the half-duplex and full-duplex energy harvesting paradigms, by generalizing the \emph{stochastic dominance technique} for RF EH-networks. We verified our analytical findings by simulating the exact and equivalent systems. The extension of our analysis to a random access network with multiple nodes can provide further insights to the development of efficient medium access protocols for networks with RF energy harvesting capabilities, and presents itself as a promising future research direction. 

\bibliographystyle{IEEEtran}
\bibliography{IEEEabrv,DiversityLib}


\end{document}